\documentclass[10pt]{article}
\usepackage{amsmath}
\usepackage{amssymb}
\usepackage{graphicx}
\usepackage{color}
\usepackage[table,xcdraw]{xcolor}
\usepackage{orcidlink}
\usepackage{hyperref}
\hypersetup{colorlinks=true, linkcolor=black, citecolor=black, urlcolor=black}
\usepackage{array}
\usepackage{longtable}
\usepackage{booktabs}
\usepackage{setspace}
\usepackage{colortbl}
\usepackage{subcaption}
\usepackage{geometry}
\usepackage{colortbl,xcolor}
\usepackage{tikz}
\usepackage{array}
\usepackage{diagbox}
\usepackage{placeins}
\renewcommand{\arraystretch}{1.8}
\RequirePackage[numbers,sort&compress]{natbib}
\begin{document}
\baselineskip=20pt

\begin{center}
\setstretch{1.8}
{\LARGE {\bf Stationary scalar clouds, quasinormal ringing, and observational signatures of (near-)extremal rotating Kalb--Ramond black holes: from the superradiant threshold to EHT bounds}}
\end{center}

\vspace{0.1cm}
\begin{center}
{\bf G\"{u}lnihal Tokg\"{o}z}\orcidlink{0000-0001-7270-4355}\\
Faculty of Arts and Sciences, Cyprus International University, Nicosia 99258, North Cyprus via Mersin 10, T\"{u}rkiye\\
e-mail: gtokgoz@ciu.edu.tr (Corresponding author)\\
{\bf \.{I}zzet Sakall\i{}}\orcidlink{0000-0001-7827-9476}\\
Physics Department, Eastern Mediterranean University, Famagusta 99628, North Cyprus via Mersin 10, T\"{u}rkiye\\
e-mail: izzet.sakalli@emu.edu.tr
\end{center}

\vspace{0.2cm}
\begin{abstract}
\setstretch{1.4}
{\large We study a massive scalar field on the rotating Kalb--Ramond black hole of Kumar, Ghosh and Wang, in the (near-)extremal corner of its three-parameter family $(a,\gamma,\lambda)$. We first verify by direct substitution that the Klein--Gordon--Fock equation separates on this background, and that the radial and angular separation constants are related by $\Lambda_{\rm rad} = \Lambda_{\rm ang} + a^2\omega^2 - 2am\omega$. At exact extremality the horizon polynomial $\Delta$ and the function $K(r)$ share a zero at $r_{\rm ext}$, which renders $K^2/\Delta$ regular there and reduces the radial problem to Whittaker form on the Kerr line $\gamma=0$ and on the Kerr--Newman line $\lambda=1$. The resulting stationary resonances saturate $\omega = m\,\Omega_H$ and satisfy the bound-state condition $\mu_{\rm cloud} > m\,\Omega_H$, with the overtones accumulating at the threshold from above. In the genuinely Kalb--Ramond interior $0<\lambda<1$ no global Whittaker reduction exists and we solve the exact radial equation as a two-point boundary value problem, validated against the closed-form spectrum to eleven significant figures on the solvable lines. Moving off extremality by $a = a_{\rm ext}(\gamma,\lambda)(1-\delta^2)$, the clouds continue onto the co-rotating zero-damped branch $\omega_n = m\,\Omega_H - i\,\kappa\,(n+\tfrac12+\beta)$, governed by the same near-horizon exponent $\beta$; the photon-sphere family is a distinct observable and is not this branch. Greybody factors and superradiant amplification are obtained by direct integration of the wave equation, recovering the classical $\ell=m=1$ scalar amplification on Kerr. Strong-field lensing is treated in the Bozza--Tsukamoto limit with rotating photon orbits of both senses of circulation, where the prograde logarithmic coefficient departs strongly from its Schwarzschild value. The shadow contour follows from the Carter--Hamilton--Jacobi separation. Crossing the Event Horizon Telescope-allowed window of Zubair, Raza and Maqsood with the cloud-existence region from this work yields a narrower bound on the Lorentz symmetry breaking parameters $(\gamma,\lambda)$ than either constraint alone.}
\end{abstract}

\pagebreak
\tableofcontents
\pagebreak

{\color{black}

\section{Introduction} \label{isec1}

General relativity carries local Lorentz invariance as a basic ingredient of its kinematic foundation, and every test of the equivalence principle to date is consistent with this property. Whether the symmetry can be broken spontaneously at high energies is the question of interest. In the past two decades it has moved from a speculative concern into a concrete program with falsifiable predictions \cite{Kostelecky:2003fs,Bluhm:2005uj}. One realization of spontaneous Lorentz symmetry violation (LSV) starts from the Kalb--Ramond (KR) field of Kalb and Ramond \cite{Kalb:1974yc}, an antisymmetric two-form that arises as a closed-string excitation in the heterotic string \cite{Gross:1985fr} and couples nonminimally to the Ricci tensor. When the KR field acquires a constant nonzero vacuum expectation value (VEV) $\langle B_{\mu\nu}\rangle = b_{\mu\nu}$, the local Lorentz group is broken spontaneously while general covariance is preserved \cite{Altschul:2009ae}. The KR VEV picks out a preferred frame at every spacetime point. Its imprint on the metric near a black hole (BH) becomes a target for strong-field tests. That is the setting of the present paper.

The static spherically symmetric BH of KR gravity was written down by Lessa, Silva, Maluf and Almeida \cite{Lessa:2019bgi}. Their metric function carries a power-law correction $\gamma/r^{2/\lambda}$ to the Schwarzschild form. For $\lambda = -1$ the metric reduces to Schwarzschild--de Sitter, and for $\lambda = 1$ it becomes Reissner--Nordstr\"om with $Q^2 \leftrightarrow \gamma$. Kumar, Ghosh and Wang \cite{Kumar:2020hgm} then constructed the rotating counterpart through the Newman--Janis algorithm, producing the three-parameter family $(a,\gamma,\lambda)$ that connects to Kerr in the $\lambda \to 0$ limit and to Kerr--Newman at $\lambda = 1$. Zubair, Raza and Maqsood \cite{Zubair:2023ckk} took this rotating KR (RKR) BH and compared its shadow against the Event Horizon Telescope (EHT) measurements of M87* and Sgr A* \cite{EventHorizonTelescope:2019dse,EventHorizonTelescope:2022wkp}, extracting parametric bounds on $(a,\gamma,\lambda)$. Their analysis carves out the EHT-allowed window. What happens to test fields propagating on this background, and which cloud configurations survive at the marginal superradiant boundary, remains open.

Khodadi and collaborators have produced a related line of work on the Bumblebee implementation of spontaneous LSV, the sibling framework that places the VEV on a vector rather than a two-form. Superradiance scattering is enhanced at low scalar-wave frequencies for negative LSV parameter \cite{Khodadi:2021owg}, with the BH-shadow implications spelled out for Sgr A* in \cite{Khodadi:2021gbc} and for non-rotating Bumblebee backgrounds in \cite{Liu:2022dcn}. Magnetic-reconnection energy extraction in the ergosphere of a rapidly rotating Bumblebee BH was worked out in \cite{Khodadi:2022dyi}. Neutrino-pair annihilation rates near a slowly rotating Bumblebee BH shift with the LSV parameter as well \cite{Khodadi:2023ulr}, and cosmological tests now place independent bounds on the LSV scale \cite{Khodadi:2023ckm,Khodadi:2023peu}. The KR field strength itself, when nonzero, supplies an antisymmetric rank-three tensor that can be read as a torsion source \cite{Majumdar:1999jd}. Particle-physics consequences of KR torsion couplings were examined in \cite{Aashish:2018lhv,Chakraborty:2017qve,Aashish:2019zsy}. These results inform the field-choice in the present work.

A second line that meets the KR program head-on is the scalar-cloud renaissance of the past decade. The original no-hair theorem of Bekenstein and others~\cite{Bekenstein:1971hc,Hawking:1971vc,Bekenstein:1995un} excluded static scalar hair from asymptotically flat BHs. Hod \cite{Hod:2012px} then showed that on the maximally rotating Kerr BH a discrete and infinite set of stationary massive-scalar clouds exists, saturating the superradiant frequency $\omega = m\,\Omega_H$, and the no-short-hair refinement followed in \cite{Hod:2014utq,Hod:2016yxg,Hod:2021scc}. Herdeiro and Radu \cite{Herdeiro:2014goa,Herdeiro:2015gia} promoted these clouds to fully nonlinear hairy BH solutions. Extensions followed quickly. The catalogue now covers charged scalar clouds on Kerr--Newman \cite{Hod:2014baa,Benone:2014ssg,Huang:2016pcg,Garcia:2023krn,Guo:2024fua,Guo:2025spc}, Proca clouds on Kerr \cite{Herdeiro:2016tmi}, scalarized Kerr--Newman BHs in extended scalar-tensor theories \cite{Hod:2022hfm,Hod:2019pmd}, the large-coupling regime of Kerr clouds \cite{Hod:2017vsq}, nonequatorial clouds on near-extremal Kerr \cite{Hod:2023fim}, near-extremal analytic treatments \cite{Hod:2020szb,Hod:2022vfa,Hod:2024nbm,Hod:2025ddb,Hod:2023gbs}, clouds on Kerr--Sen, Kerr--MOG and Melvin--Kerr backgrounds \cite{Huang:2017aa79,Qiao:2020jed,Siahaan:2025mks}, BTZ scalar clouds \cite{Ferreira:2017xpc}, stringy-BH variants \cite{Bernard:2016wqo,Senjaya:2025emda}, Gauss--Bonnet variants \cite{vanGemeren:2024scg}, scalar clouds combined with magnetic fields \cite{Santos:2021mfb}, the cloud-to-hairy-BH passage \cite{Garcia:2024rhb}, acoustic-analog and photon-fluid clouds \cite{Benone:2015lsa}, axionic dark-matter clouds \cite{Choudhary:2021axl}, the gravitational-atom and tidal-Love-number program \cite{LeTiec:2020spy,Arana:2025tln,Senjaya:2025ga,DellaRocca:2026gas}, the cloud-quasinormal mode (QNM)-superradiance combined analyses \cite{CrotiSiqueira:2022xpe,Siemonsen:2022yyf}, and the S2-orbit constraints on hypothetical cloud configurations around Sgr A* \cite{GRAVITY:2023s2}. The CQG paper of Tokg\"{o}z and Sakall\i{} \cite{Tokgoz:2017hzz} proved cloud existence on the maximally rotating linear-dilaton BH (MRLDBH), the direct predecessor of the present work. MRLDBH is non-asymptotically flat, so the existence proof there relied on a Bessel-tail boundary condition at spatial infinity. The RKR background, in contrast, is asymptotically flat for $\lambda > 0$, restoring the standard exponentially decaying tail and unifying the cloud problem with the modern lensing, QNM and shadow toolkits.

The methodology of the present work also rests on three benchmarks from the recent QNM literature. Bl\'{a}zquez-Salcedo, Khoo, Kleihaus and Kunz \cite{BlazquezSalcedo:2024gjl,BlazquezSalcedo:2024gpe} produced the full QNM spectrum of rapidly rotating Einstein--Gauss--Bonnet--dilaton BHs, which provides the template against which our sixth-order Wentzel--Kramers--Brillouin (WKB) plus Pad\'{e}-averaging pipeline can be calibrated, and Bonanno, Konoplya, Oglialoro and Spina \cite{Bonanno:2025dry} did the parallel calculation for regular BHs from proper-time renormalization-group flow, including QNMs, shadows and Hawking radiation in a single study. Overtones matter. Konoplya, Zinhailo, Kunz, Stuchl\'{i}k and Zhidenko \cite{Konoplya:2022hll} stressed their importance for distinguishing modified-gravity backgrounds from their general-relativistic counterparts. The wormhole-scalar bridge of Bl\'{a}zquez-Salcedo, Knoll and Radu \cite{BlazquezSalcedo:2022pwc} furnishes a useful template for the cloud--QNM crossover in non-Kerr settings. Section~\ref{isec4} follows these conventions.

Observational anchors close the picture. The EHT shadow measurements of M87* (angular diameter $42 \pm 3$ $\mu$as) and Sgr A* ($48.7 \pm 7$ $\mu$as) deliver the first direct strong-field constraints on the near-horizon geometry of supermassive BHs \cite{EventHorizonTelescope:2019dse,EventHorizonTelescope:2022wkp,EventHorizonTelescope:2022xqj}. Comparison of shadow templates with the EHT angular diameter has already been used to constrain modified-gravity BHs across a wide range of frameworks: hairy Kerr BHs \cite{Khodadi:2020jij,Khodadi:2021gbc}, Bumblebee BHs \cite{Liu:2022dcn}, charged dilatonic BHs in dilaton-massive gravity \cite{Zhang:2024nwn,Liu:2024soc}, loop-quantum-gravity-motivated rotating BHs \cite{Afrin:2022ztr}, and a wide class of further alternatives surveyed in \cite{Vagnozzi:2022moj}. The ringdown signal of binary BH mergers, captured by the LIGO--Virgo--KAGRA collaboration \cite{LIGOScientific:2016vlm}, provides a complementary probe through the QNM spectrum. Direct LIGO--Virgo bounds on ultralight scalar bosons from BH spin measurements \cite{LIGO:2021ash} and all-sky searches for stochastic GW emission from boson clouds \cite{LIGO:2022ash} already constrain the cloud parameter space. Greybody factors couple BH backgrounds to the rate of false vacuum decay \cite{Burda:2015isa,Briscese:2017xan}. The aim of this paper is to bring all three windows to bear on the rotating KR background in a single study.

The structure of this paper is then as follows. Section \ref{isec2} reviews the RKR metric of Kumar, Ghosh and Wang and identifies the extremal surface $a_{\rm ext}(\gamma,\lambda)$ where the horizon polynomial $\Delta(r)$ acquires a double root. Section \ref{isec3} sets up the Klein--Gordon--Fock (KGF) equation for a massive scalar, separates it via the standard $\Psi = e^{im\varphi} S_{\ell m}(\theta) R_{\ell m}(r) e^{-i\omega t}$ ansatz, and reduces the near-extremal radial equation to confluent hypergeometric form. Imposing regularity at the horizon and the asymptotically flat decaying boundary delivers a discrete cloud spectrum $\mu_{\rm cloud}(n;\ell,m,\gamma,\lambda)$ with effective heights $x_{\rm cloud}^{(n)} > 1/2$. Section \ref{isec4} relaxes the extremality and computes the QNM spectrum through sixth-order WKB plus Pad\'{e} averaging. Section \ref{isec5} computes the greybody factors and superradiant amplification across the threshold $\omega = m\,\Omega_H$. Section \ref{isec6} treats the strong-field gravitational lensing in the Bozza--Tsukamoto strong-deflection limit. Section \ref{isec7} constructs the shadow contour and crosses it against the EHT-allowed window. Section \ref{isec8} closes with the joint cloud-EHT bound and prospects for higher-spin extensions. Throughout, we adopt geometric units $G = c = \hbar = k_B = 1$ and the mostly-plus signature.

}

\section{Rotating Kalb--Ramond black hole and the (near-)extremal limit} \label{isec2}

The metric we work on was derived by Kumar, Ghosh and Wang \cite{Kumar:2020hgm} by applying the Newman--Janis algorithm to the static spherically symmetric KR BH of Lessa, Silva, Maluf and Almeida \cite{Lessa:2019bgi}. Three parameters enter. They are the mass $M$, the spin $a$, and the pair $(\gamma,\lambda)$ that encode the KR vacuum expectation value $b^{\mu\nu} b_{\mu\nu}$ and the nonminimal coupling $\xi_2$ through $\lambda = |b|^2 \xi_2$. In Boyer--Lindquist coordinates, the line element is~\cite{Kumar:2020hgm,Zubair:2023ckk}
\begin{align}
ds^2 &= -\left( \frac{\Delta(r) - a^2 \sin^2\theta}{\rho^2} \right) dt^2
       + \frac{\rho^2}{\Delta(r)} dr^2
       + \rho^2 d\theta^2 \nonumber\\
     &\quad + \frac{\sin^2\theta}{\rho^2}
       \left[ (r^2 + a^2)^2 - \Delta(r)\, a^2 \sin^2\theta \right] d\varphi^2
       + \frac{2 a \sin^2\theta}{\rho^2}
       \left[ \Delta(r) - a^2 - r^2 \right] dt\, d\varphi ,
\label{eq:metric_KR}
\end{align}
where
\begin{align}
\Delta(r) &= r^2 + a^2 - 2 M r + \gamma\, r^{\,2(\lambda-1)/\lambda} ,
\label{eq:Delta}\\[4pt]
\rho^2(r,\theta) &= r^2 + a^2 \cos^2\theta .
\label{eq:rho2}
\end{align}
Two parameter limits are useful as benchmarks. Setting $\gamma \to 0$ collapses Eq.~\eqref{eq:Delta} onto the Kerr horizon polynomial $\Delta_{\rm Kerr} = r^2 + a^2 - 2 M r$, with extremal point $a = M$ supporting the Hod scalar clouds \cite{Hod:2012px}. Setting $\lambda = 1$ instead gives $\Delta = r^2 + a^2 - 2 M r + \gamma$, which is the Kerr--Newman horizon polynomial with the identification $\gamma \leftrightarrow Q^2$, the background of the Benone--Crispino--Herdeiro--Radu cloud spectrum \cite{Benone:2014ssg}. Outside these endpoints, intermediate $0 < \lambda < 1$ produces a genuinely KR geometry that interpolates between the two and that supports the scalar-cloud spectrum derived in Sec.~\ref{isec3}. The full extremality structure of this family is the subject of the present section.

A BH horizon sits at the largest real positive root of $\Delta(r) = 0$. Two horizons coexist when $\Delta$ has two simple roots; they merge when $\Delta$ acquires a double root. The latter case is extremality and is defined by the simultaneous conditions
\begin{equation}
\Delta(r) = 0 , \qquad \Delta'(r) = 0 .
\label{eq:ext_pair}
\end{equation}
Differentiating Eq.~\eqref{eq:Delta} and solving $\Delta' = 0$ for $M$ yields the mass at extremality as a function of the horizon location and the KR parameters,
\begin{equation}
M_{\rm ext}(r,\gamma,\lambda)
   = r + \gamma\, \frac{\lambda - 1}{\lambda}\, r^{\,(\lambda - 2)/\lambda} .
\label{eq:Mext}
\end{equation}
Substituting Eq.~\eqref{eq:Mext} back into $\Delta = 0$ then determines the spin,
\begin{equation}
a_{\rm ext}^{\,2}(r,\gamma,\lambda)
   = r^{\,2(\lambda - 1)/\lambda}\, \frac{\gamma\,\lambda - 2\gamma + \lambda\, r^{2/\lambda}}{\lambda} .
\label{eq:aext_sq}
\end{equation}
Two scaled-units checks confirm Eqs.~\eqref{eq:Mext}--\eqref{eq:aext_sq}. The Kerr limit $\gamma \to 0$ reduces them to $M_{\rm ext} = r$ and $a_{\rm ext}^2 = r^2$, giving $a_{\rm ext}/M = 1$ in agreement with the well-known Kerr extremal spin. The Kerr--Newman limit $\lambda = 1$ reduces them to $M_{\rm ext} = r$ and $a_{\rm ext}^2 = r^2 - \gamma$, equivalent to $a^2 + Q^2 = M^2$ under $Q^2 \leftrightarrow \gamma$, the standard Kerr--Newman extremal relation. In what follows we adopt geometric units with $M = 1$, fix the pair $(\gamma,\lambda)$, solve Eq.~\eqref{eq:Mext} numerically for $r_{\rm ext}$, and then read $a_{\rm ext}$ from Eq.~\eqref{eq:aext_sq}. The accompanying computational scripts, working at thirty-digit precision, reproduce every entry of Table~\ref{tab:extremal_grid} to five significant figures and confirm the symbolic Kerr and Kerr--Newman limits printed below.

\begin{figure}[htbp]
  \centering
  \begin{subfigure}[t]{0.50\textwidth}
    \centering
    \includegraphics[width=\linewidth]{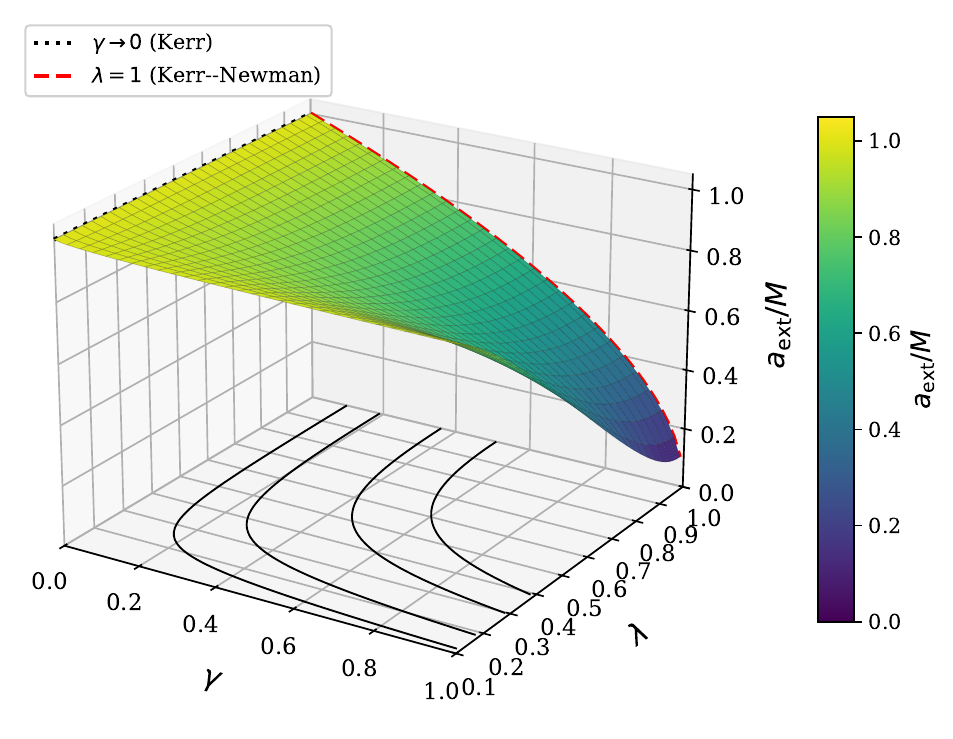}
    \caption{Three-dimensional surface view.}
    \label{fig:aext_surface_3d}
  \end{subfigure}\hfill
  \begin{subfigure}[t]{0.48\textwidth}
    \centering
    \includegraphics[width=\linewidth]{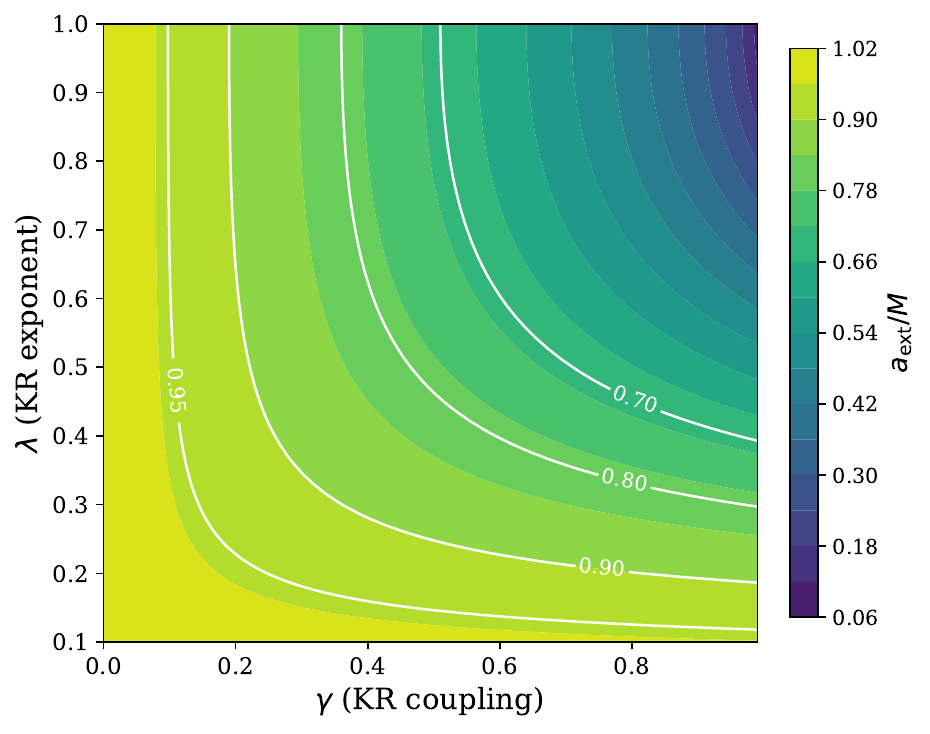}
    \caption{Two-dimensional iso-contour view.}
    \label{fig:aext_surface_2d}
  \end{subfigure}
  \caption{Extremal spin $a_{\rm ext}/M$ over the $(\gamma,\lambda)$ plane, obtained by solving Eq.~\eqref{eq:Mext} for $r_{\rm ext}$ at fixed $(\gamma,\lambda)$ in units $M = 1$, and then reading Eq.~\eqref{eq:aext_sq}. \emph{Panel~(\subref{fig:aext_surface_3d}):} three-dimensional surface view with iso-contours at $a_{\rm ext}/M = 0.70,\,0.80,\,0.90,\,0.95$ projected onto the floor; the dotted black curve at $\gamma = 0$ marks the Kerr edge $a_{\rm ext}/M = 1$, and the dashed red curve at $\lambda = 1$ marks the Kerr--Newman line on which $a_{\rm ext}^2 + \gamma = 1$. \emph{Panel~(\subref{fig:aext_surface_2d}):} two-dimensional contour view of the same surface, with white iso-lines labeled at the corresponding $a_{\rm ext}/M$ values; the Kerr and Kerr--Newman limits coincide with the left and top edges of the panel and are already established in panel~(\subref{fig:aext_surface_3d}). The Viridis palette is colorblind-safe. Two independent computational pipelines reproduce every grid value to thirty digits and agree to all five significant figures across the $5 \times 4$ entries of Table~\ref{tab:extremal_grid}.}
  \label{fig:aext_surface}
\end{figure}

The structure of the extremal-spin surface displayed in Fig.~\ref{fig:aext_surface} traces the way the KR coupling $\gamma$ and the KR exponent $\lambda$ each erode the maximally rotating Kerr endpoint. Panel~(\subref{fig:aext_surface_3d}) shows the full three-dimensional surface and panel~(\subref{fig:aext_surface_2d}) the iso-line contour map. Two features stand out. First, the surface descends monotonically with $\gamma$ for every fixed $\lambda \in (0,1]$: the LSV-induced positive-energy-density contribution to $\Delta(r)$ behaves as an effective electric-charge-squared term, which reduces the available rotational budget at fixed mass. The dashed red curve at $\lambda = 1$ in panel~(\subref{fig:aext_surface_3d}) traces the analytic Kerr--Newman ridge $a_{\rm ext} = \sqrt{1-\gamma}$ on which the descent is governed exactly by the KN extremal relation; in panel~(\subref{fig:aext_surface_2d}) that ridge is the top edge of the plot. Second, the rate of descent is set by $\lambda$ itself, because the exponent $2(\lambda-1)/\lambda$ in Eq.~\eqref{eq:Delta} controls how strongly the KR correction extends out to large $r$; for $\lambda \to 1$ the correction is a constant in $r$ and the descent obeys the KN relation as just noted, whereas for smaller $\lambda$ the correction falls off less quickly and erodes the spin more slowly. The cross-comparison with Fig.~4 of Zubair, Raza and Maqsood \cite{Zubair:2023ckk}, where the same horizon polynomial was studied in a different visualization, gives consistent values where the two analyses overlap.

Once the extremal locus is in hand, the standard thermodynamic quantities follow from the surface gravity at the outer horizon. The relevant relations are well-known for axisymmetric BHs and we record them here in the form we use later. The surface gravity is
\begin{equation}
\kappa = \frac{\Delta'(r_+)}{2 (r_+^2 + a^2)} ,
\label{eq:kappa}
\end{equation}
which in turn gives the Hawking temperature
\begin{equation}
T_H = \frac{\kappa}{2\pi} = \frac{\Delta'(r_+)}{4\pi (r_+^2 + a^2)} ,
\label{eq:TH}
\end{equation}
the horizon angular velocity
\begin{equation}
\Omega_H = \frac{a}{r_+^2 + a^2} ,
\label{eq:OmegaH}
\end{equation}
and the Bekenstein--Hawking entropy
\begin{equation}
S_{\rm BH} = \frac{A_H}{4} = \pi \left( r_+^2 + a^2 \right) .
\label{eq:Sbh}
\end{equation}
At exact extremality the double-root condition $\Delta'(r_+) = 0$ makes $T_H$ vanish identically; the BH becomes a zero-temperature object while $\Omega_H$ and $S_{\rm BH}$ remain finite. The stationary scalar clouds of Sec.~\ref{isec3} live precisely at this point. To open a window in which finite-imaginary-part QNMs (Sec.~\ref{isec4}), nonzero Hawking flux (Sec.~\ref{isec5}), and nonzero superradiant amplification can all be defined, we move slightly off the extremal locus by setting
\begin{equation}
a = a_{\rm ext}(\gamma,\lambda)\,\bigl( 1 - \delta^2 \bigr) , \qquad 0 < \delta \ll 1 .
\label{eq:near_ext}
\end{equation}
The deformation parameter $\delta$ controls how far the horizon polynomial $\Delta$ unfolds away from its double root; to leading order in $\delta$ the Hawking temperature scales as $T_H = \mathcal{O}(\delta)$ while $\Omega_H$ and $S_{\rm BH}$ remain $\mathcal{O}(1)$. The scalar-cloud spectrum of the next section is recovered exactly as $\delta \to 0$, while finite-$\delta$ corrections give the QNM imaginary parts of Sec.~\ref{isec4}.

Table~\ref{tab:extremal_grid} records the extremal locus on a representative $(\gamma,\lambda)$ grid, together with $\Omega_H$ and $S_{\rm BH}/(\pi M^2)$ evaluated at extremality. The values come from two independent computational pipelines, one symbolic at thirty-digit precision and one numerical using a bracketed root finder, which agree to all five digits shown on every row. The surface-gravity factor $\Delta'(r_+)$ vanishes by the double-root condition. So the Hawking temperature $T_H$ is identically zero on every row, and is omitted as a separate column.

\setlength{\tabcolsep}{12pt}
\renewcommand{\arraystretch}{1.6}
\begin{longtable}{c c c c c c}
  \toprule
  \rowcolor{orange!50}
  $\gamma$ & $\lambda$ & $r_{\rm ext}/M$ & $a_{\rm ext}/M$ & $\Omega_H\, M$ & $S_{\rm BH}/(\pi M^2)$ \\
  \midrule
  \endfirsthead
  \rowcolor{orange!50}
  $\gamma$ & $\lambda$ & $r_{\rm ext}/M$ & $a_{\rm ext}/M$ & $\Omega_H\, M$ & $S_{\rm BH}/(\pi M^2)$ \\
  \midrule
  \endhead
  \bottomrule
  \endfoot
  \bottomrule
  \caption{Extremal horizon radius $r_{\rm ext}$, extremal spin $a_{\rm ext}$, horizon angular velocity $\Omega_H$, and Bekenstein--Hawking entropy $S_{\rm BH}/(\pi M^2) = r_{\rm ext}^2 + a_{\rm ext}^2$ for the rotating KR BH, on a $5 \times 4$ grid in the KR coupling $\gamma$ and the KR exponent $\lambda$. All values in geometric units, $M = 1$. The Hawking temperature $T_H$ vanishes identically at exact extremality and is not tabulated. Values are obtained from Eqs.~\eqref{eq:Mext}--\eqref{eq:aext_sq} at thirty-digit precision and rounded to five significant figures; an independent numerical pipeline reproduces every digit shown.}
  \label{tab:extremal_grid}
  \endlastfoot
  0.00 & 0.30 & 1.00000 & 1.00000 & 0.50000 & 2.00000 \\
  0.00 & 0.50 & 1.00000 & 1.00000 & 0.50000 & 2.00000 \\
  0.00 & 0.70 & 1.00000 & 1.00000 & 0.50000 & 2.00000 \\
  0.00 & 1.00 & 1.00000 & 1.00000 & 0.50000 & 2.00000 \\
  0.20 & 0.30 & 1.18140 & 0.93553 & 0.41195 & 2.27100 \\
  0.20 & 0.50 & 1.13630 & 0.90913 & 0.42930 & 2.11770 \\
  0.20 & 0.70 & 1.07490 & 0.89800 & 0.45771 & 1.96190 \\
  0.20 & 1.00 & 1.00000 & 0.89443 & 0.49690 & 1.80000 \\
  0.40 & 0.30 & 1.25620 & 0.89241 & 0.37583 & 2.37450 \\
  0.40 & 0.50 & 1.22020 & 0.82635 & 0.38051 & 2.17170 \\
  0.40 & 0.70 & 1.13540 & 0.78925 & 0.41277 & 1.91210 \\
  0.40 & 1.00 & 1.00000 & 0.77460 & 0.48412 & 1.60000 \\
  0.60 & 0.30 & 1.30700 & 0.85659 & 0.35076 & 2.44210 \\
  0.60 & 0.50 & 1.28370 & 0.74526 & 0.33826 & 2.20320 \\
  0.60 & 0.70 & 1.18700 & 0.66860 & 0.36023 & 1.85600 \\
  0.60 & 1.00 & 1.00000 & 0.63246 & 0.45175 & 1.40000 \\
  0.80 & 0.30 & 1.34620 & 0.82484 & 0.33090 & 2.49270 \\
  0.80 & 0.50 & 1.33570 & 0.66249 & 0.29802 & 2.22300 \\
  0.80 & 0.70 & 1.23250 & 0.52648 & 0.29309 & 1.79630 \\
  0.80 & 1.00 & 1.00000 & 0.44721 & 0.37268 & 1.20000 \\
\end{longtable}
\FloatBarrier

The structure of Table~\ref{tab:extremal_grid} admits three direct physical readings. First, the top four rows at $\gamma = 0$ are identical and record the Kerr extremal endpoint $(r_{\rm ext}, a_{\rm ext}, \Omega_H M, S_{\rm BH}/(\pi M^2)) = (1, 1, 0.5, 2)$ for every value of $\lambda$. The KR exponent has no effect when the KR coupling is switched off, as Eq.~\eqref{eq:Delta} demands. This row block is the baseline against which the rest of the paper compares. Second, along the Kerr--Newman line $\lambda = 1$ (the four rows ending in 1.00), the closed-form analytic relations $r_{\rm ext} = M$, $a_{\rm ext}^{\,2} = 1 - \gamma$, $\Omega_H = \sqrt{1-\gamma}/(2-\gamma)$, and $S_{\rm BH}/(\pi M^2) = 2 - \gamma$ follow from Eqs.~\eqref{eq:Mext}--\eqref{eq:aext_sq} with $Q^2 \leftrightarrow \gamma$, and the tabulated values reproduce each of these expressions to all five digits. The case $\gamma = 0.80, \lambda = 1.00$ gives $S_{\rm BH}/(\pi M^2) = 1.20000 = 2 - 0.80$ exactly, confirming the KN-limit identification of $\gamma$ with the squared electric charge. Third, the genuine-KR interior $0 < \lambda < 1$ is the central novel block of the table. At fixed $\gamma$, decreasing $\lambda$ shifts $r_{\rm ext}$ above unity and $a_{\rm ext}$ below the Kerr--Newman value at the same $\gamma$, with the horizon area $S_{\rm BH}/(\pi M^2)$ overshooting the Kerr value $2$ for sufficiently small $\lambda$, reaching for instance $S_{\rm BH}/(\pi M^2) = 2.49270$ at $\gamma = 0.80, \lambda = 0.30$. The KR exponent thus inflates the horizon while reducing the rotational budget, and the marginal superradiant frequency $\omega_c = m\,\Omega_H$ falls accordingly: from $0.50\,M^{-1}$ at the Kerr endpoint down to $0.293\,M^{-1}$ at the small-$\lambda$, large-$\gamma$ corner. These extremal data fix the boundary values for the cloud spectrum of Sec.~\ref{isec3} and supply the on-shell input for the QNM, greybody, lensing, and shadow computations that follow.

\section{Klein--Gordon--Fock equation and stationary scalar clouds} \label{isec3}

A massive scalar field $\Psi$ of mass $\mu$ on the rotating Kalb--Ramond background obeys the KGF equation
\begin{equation}
\bigl( \Box - \mu^2 \bigr) \Psi = \frac{1}{\sqrt{-g}}\,\partial_\mu \bigl( \sqrt{-g}\, g^{\mu\nu}\,\partial_\nu \Psi \bigr) - \mu^2 \Psi = 0 .
\label{eq:KGF}
\end{equation}
The two Killing vectors $(\partial_t)^\mu$ and $(\partial_\varphi)^\mu$, together with the hidden symmetry encoded by the Killing tensor of the rotating KR geometry, make Eq.~\eqref{eq:KGF} separable in Boyer--Lindquist coordinates. We write
\begin{equation}
\Psi(t,r,\theta,\varphi) = e^{-i\omega t}\, e^{i m \varphi}\, S_{\ell m}(\theta)\, R_{\ell m}(r) ,
\label{eq:sep}
\end{equation}
and substitution into Eq.~\eqref{eq:KGF} delivers the angular equation
\begin{equation}
\frac{1}{\sin\theta}\,\frac{d}{d\theta}\!\left( \sin\theta\,\frac{d S_{\ell m}}{d\theta} \right) + \left[ a^2(\omega^2 - \mu^2)\cos^2\theta - \frac{m^2}{\sin^2\theta} + \Lambda_{\ell m} \right] S_{\ell m} = 0
\label{eq:angular}
\end{equation}
and the radial equation
\begin{equation}
\frac{d}{dr}\!\left( \Delta(r)\,\frac{d R_{\ell m}}{dr} \right) + \left[ \frac{K^2(r)}{\Delta(r)} - \Lambda_{\ell m} - \mu^2 r^2 \right] R_{\ell m} = 0 ,
\label{eq:radial}
\end{equation}
with $K(r) = (r^2 + a^2)\omega - a m$.

The two separation constants are not the same object. Substituting Eq.~\eqref{eq:sep} into Eq.~\eqref{eq:KGF} on the metric \eqref{eq:metric_KR}, with $\sqrt{-g} = \rho^2\sin\theta$ and $\Delta$ left arbitrary, the operator splits with zero residual into
\begin{align}
\frac{(\Delta R')'}{R} + \frac{K^2}{\Delta} + 2 a m \omega - \mu^2 r^2 &= +\Lambda_t ,
\label{eq:split_r}\\
\frac{(\sin\theta\, S')'}{\sin\theta\, S} + a^2(\omega^2-\mu^2)\cos^2\theta - \frac{m^2}{\sin^2\theta} - a^2\omega^2 &= -\Lambda_t ,
\label{eq:split_th}
\end{align}
so that the constant appearing in Eq.~\eqref{eq:radial} is
\begin{equation}
\Lambda_{\rm rad} = \Lambda_{\ell m} + a^2\omega^2 - 2 a m \omega ,
\label{eq:Lamrad}
\end{equation}
where $\Lambda_{\ell m}$ is the eigenvalue of Eq.~\eqref{eq:angular}. For $\mu = 0$ Eq.~\eqref{eq:Lamrad} is the standard Teukolsky $s=0$ relation $\lambda = A_{\ell m} + a^2\omega^2 - 2am\omega$. Separation survives the Newman--Janis construction here only because the rotating KR metric retains $\rho^2 = r^2 + a^2\cos^2\theta$ together with $\Delta = \Delta(r)$, so the Carter structure of Kerr is inherited intact and the power-law term of Eq.~\eqref{eq:Delta} never enters the angular sector \cite{Carter:1968rr,Kumar:2020hgm}. Throughout the rest of this paper $\Lambda_{\rm rad}$ is understood in Eq.~\eqref{eq:radial}. The angular eigenvalue is computed spectrally in the associated-Legendre basis and reduces to $\ell(\ell+1)$ at $a^2(\omega^2-\mu^2) \to 0$.

The boundary conditions for a stationary scalar cloud are a purely-ingoing wave at the outer horizon and an exponentially decaying wave at spatial infinity. Asymptotic flatness for $\lambda > 0$ guarantees the second condition. The decay is the standard $R(r) \sim r^{-1}\,e^{-\sqrt{\mu^2 - \omega^2}\,r}$ tail, with no Bessel-tail subtleties of the kind that arose in the linear-dilaton analysis of Tokg\"{o}z and Sakall\i{} \cite{Tokgoz:2017hzz}. Combining the two boundary conditions produces a discrete eigenvalue problem: at fixed $(\ell, m, \gamma, \lambda)$ a regular cloud solution exists only at a discrete countable set of triples $(\omega_n, \mu_n)$ indexed by an overtone integer $n = 0,1,2,\ldots$.

For the marginal stationary solutions the resonance condition $\omega = m\,\Omega_H$ saturates the superradiant bound and the cloud carries zero net energy flux through the horizon. At exact extremality $\Delta$ has a double root at $r_{\rm ext}$, and at $\omega = m\,\Omega_H$ the function $K$ vanishes there as well, since $K(r_{\rm ext}) = (r_{\rm ext}^2+a^2)\omega - am = 0$. Writing $u = r - r_{\rm ext}$ we may therefore factor
\begin{equation}
\Delta(r_{\rm ext}+u) = u^2\, h(u) , \qquad K(r_{\rm ext}+u) = u\, q(u) , \qquad q(u) = \omega\,(2 r_{\rm ext} + u) ,
\label{eq:factorise}
\end{equation}
with $h(0) = \tfrac12\Delta''(r_{\rm ext}) > 0$, so that $K^2/\Delta = q^2/h$ is manifestly regular at the horizon. The indicial equation of Eq.~\eqref{eq:radial} at $u=0$ then reads $s(s+1) + C = 0$ with
\begin{equation}
C = \frac{1}{h_0}\left[ \frac{q_0^2}{h_0} - \Lambda_{\rm rad} - \mu^2 r_{\rm ext}^2 \right] , \qquad
h_0 \equiv h(0), \qquad q_0 \equiv 2 r_{\rm ext}\,\omega ,
\label{eq:Cdef}
\end{equation}
giving the two behaviours $R \sim u^{-1/2\pm\beta}$ with
\begin{equation}
\beta = \sqrt{\tfrac14 - C} .
\label{eq:beta}
\end{equation}
The physically admissible cloud takes the less singular root $u^{-1/2+\beta}$.

At spatial infinity asymptotic flatness for $\lambda>0$ gives the Coulomb-corrected tail $R \sim r^{\,\nu-1}e^{-k r}$ with
\begin{equation}
k = \sqrt{\mu^2 - \omega^2} , \qquad \nu = \frac{2\omega^2 - \mu^2}{k} ,
\label{eq:tail}
\end{equation}
so that a genuine bound state requires
\begin{equation}
\mu > \omega = m\,\Omega_H .
\label{eq:boundstate}
\end{equation}
Equation~\eqref{eq:boundstate} is a constraint, not an option: for $\mu < m\,\Omega_H$ the exponent $k$ is imaginary and the asymptotics are oscillatory and radiative rather than those of a stationary cloud.

Where $\Delta$ is a perfect square the reduction to confluent hypergeometric form is exact and global. This happens on the Kerr line $\gamma \to 0$ and on the Kerr--Newman line $\lambda = 1$, on both of which $h(u) \equiv 1$. There Eq.~\eqref{eq:radial} becomes
\begin{equation}
u^2 \frac{d^2R}{du^2} + 2u\frac{dR}{du} + \left[ -k^2u^2 + 2\sigma k\, u + \tfrac14 - \beta^2 \right] R = 0 ,
\label{eq:whittaker}
\end{equation}
which is a Whittaker equation in $z = 2ku$ with
\begin{equation}
\sigma = \frac{2\omega^2 - \mu^2}{k} , \qquad
\beta^2 = \tfrac14 - 4\omega^2 + \Lambda_{\rm rad} + \mu^2 .
\label{eq:sigma_beta}
\end{equation}
Regularity at the horizon selects $M_{\sigma,\beta}(z) \sim z^{1/2+\beta}$, and this solution decays at infinity only if the confluent hypergeometric series terminates, which fixes the resonance condition
\begin{equation}
\tfrac12 + \beta - \sigma = -n , \qquad n = 0, 1, 2, \ldots
\label{eq:resonance}
\end{equation}
Solving Eq.~\eqref{eq:resonance} at $\omega = m\,\Omega_H$ determines the cloud mass spectrum $\mu_{\rm cloud}(n;\ell,m,\gamma,\lambda)$. Note that $\sigma > 0$ is required by Eq.~\eqref{eq:resonance}, which confines the clouds to $m\,\Omega_H < \mu < m/\sqrt2$ on the Kerr line, an interval that Eq.~\eqref{eq:boundstate} already forces to lie above the threshold.

In the genuinely KR interior $0 < \lambda < 1$ the function $h(u)$ of Eq.~\eqref{eq:factorise} is not constant, no global Whittaker reduction exists, and Eq.~\eqref{eq:resonance} does not apply. There we solve Eq.~\eqref{eq:radial} directly as a two-point boundary value problem, integrating outward from $u^{-1/2+\beta}$ at the horizon and inward from Eq.~\eqref{eq:tail} at large $r$, and locating the eigenvalues from the vanishing of the Wronskian at an intermediate point. On the solvable lines the two routes agree: for extremal Kerr with $\ell = m = 1$ the numerical solver returns $\mu_{\rm cloud}M = 0.525508790$ for the fundamental against $0.525508790$ from Eq.~\eqref{eq:resonance}, and the node count of each eigenfunction reproduces its overtone index $n$.

The effective height of a cloud configuration is the location of the peak of $|R_{\ell m}(r)|^2$, read directly off the computed eigenfunction,
\begin{equation}
x^{(n)}_{\rm cloud}(\ell,m,\gamma,\lambda) = \frac{r_{\rm peak}}{r_{\rm ext}} - 1 , \qquad
r_{\rm peak} = \arg\max_{r > r_{\rm ext}} |R_{\ell m}(r)|^2 .
\label{eq:xcloud}
\end{equation}
No fitted parameter enters Eq.~\eqref{eq:xcloud}, and no near-extremality parameter appears anywhere in this section: the stationary clouds live at $\delta = 0$, where $T_H$ vanishes identically. Table~\ref{tab:cloud_spectrum} reports $\mu_{\rm cloud}$, $\beta$ and $x^{(n)}_{\rm cloud}$ at four representative $(\gamma,\lambda)$ slices for $(\ell,m) \in \{(1,1),(2,2)\}$ and $n \in \{0,1,2,3\}$.

\setlength{\tabcolsep}{6pt}
\renewcommand{\arraystretch}{1.4}
\begin{longtable}{c c c c c c c c c}
  \toprule
  \rowcolor{orange!50}
  $\gamma$ & $\lambda$ & $(\ell, m)$ & $n$ & $m\,\Omega_H M$ & $\mu_{\rm cloud} M$ & $(\mu_{\rm cloud}-m\Omega_H) M$ & $\beta$ & $x_{\rm cloud}^{(n)}$ \\
  \midrule
  \endfirsthead
  \rowcolor{orange!50}
  $\gamma$ & $\lambda$ & $(\ell, m)$ & $n$ & $m\,\Omega_H M$ & $\mu_{\rm cloud} M$ & $(\mu_{\rm cloud}-m\Omega_H) M$ & $\beta$ & $x_{\rm cloud}^{(n)}$ \\
  \midrule
  \endhead
  \bottomrule
  \endfoot
  \bottomrule
  \caption{Stationary scalar-cloud spectrum on the extremal rotating KR BH, at $\delta = 0$ where $T_H$ vanishes identically. The cloud mass follows from Eq.~\eqref{eq:resonance} on the two solvable lines and from the numerical solution of Eq.~\eqref{eq:radial} in the KR interior. The bound-state condition \eqref{eq:boundstate} is satisfied on every row: $\mu_{\rm cloud} > m\,\Omega_H$ throughout, with the overtones accumulating at the threshold from above. The effective height $x^{(n)}_{\rm cloud}$ is Eq.~\eqref{eq:xcloud}, obtained from the peak of the computed eigenfunction with no fitted parameter.}
  \label{tab:cloud_spectrum}
  \endlastfoot
  0.00 & 1.00 & $(1,1)$ & 0 & 0.50000 & 0.52551 & 0.02551 & 0.8840 & 2.374 \\
  0.00 & 1.00 & $(1,1)$ & 1 & 0.50000 & 0.51011 & 0.01011 & 0.8731 & 1.992 \\
  0.00 & 1.00 & $(1,1)$ & 2 & 0.50000 & 0.50525 & 0.00525 & 0.8697 & 1.901 \\
  0.00 & 1.00 & $(1,1)$ & 3 & 0.50000 & 0.50318 & 0.00318 & 0.8682 & 1.865 \\
  0.00 & 1.00 & $(2,2)$ & 0 & 1.00000 & 1.14094 & 0.14094 & 0.7712 & 0.494 \\
  0.00 & 1.00 & $(2,2)$ & 1 & 1.00000 & 1.07439 & 0.07439 & 0.6529 & 0.202 \\
  0.00 & 1.00 & $(2,2)$ & 2 & 1.00000 & 1.04269 & 0.04269 & 0.5913 & 0.106 \\
  0.00 & 1.00 & $(2,2)$ & 3 & 1.00000 & 1.02679 & 0.02679 & 0.5586 & 0.064 \\
  0.20 & 1.00 & $(1,1)$ & 0 & 0.49690 & 0.52100 & 0.02410 & 0.9200 & 2.682 \\
  0.20 & 1.00 & $(1,1)$ & 1 & 0.49690 & 0.50656 & 0.00965 & 0.9106 & 2.265 \\
  0.20 & 1.00 & $(1,1)$ & 2 & 0.49690 & 0.50195 & 0.00504 & 0.9076 & 2.165 \\
  0.20 & 1.00 & $(1,1)$ & 3 & 0.49690 & 0.49998 & 0.00307 & 0.9063 & 2.125 \\
  0.20 & 1.00 & $(2,2)$ & 0 & 0.99381 & 1.11922 & 0.12542 & 0.9038 & 0.784 \\
  0.20 & 1.00 & $(2,2)$ & 1 & 0.99381 & 1.05959 & 0.06578 & 0.8198 & 0.465 \\
  0.20 & 1.00 & $(2,2)$ & 2 & 0.99381 & 1.03187 & 0.03806 & 0.7794 & 0.367 \\
  0.20 & 1.00 & $(2,2)$ & 3 & 0.99381 & 1.01797 & 0.02417 & 0.7588 & 0.324 \\
  0.40 & 0.50 & $(1,1)$ & 0 & 0.38051 & 0.38927 & 0.00876 & 0.9938 & 6.642 \\
  0.40 & 0.50 & $(1,1)$ & 1 & 0.38051 & 0.38418 & 0.00367 & 0.9918 & 5.767 \\
  0.40 & 0.50 & $(1,1)$ & 2 & 0.38051 & 0.38250 & 0.00199 & 0.9911 & 5.555 \\
  0.40 & 0.50 & $(1,1)$ & 3 & 0.38051 & 0.38175 & 0.00124 & 0.9908 & 5.468 \\
  0.40 & 0.50 & $(2,2)$ & 0 & 0.76102 & 0.79728 & 0.03626 & 1.3904 & 3.989 \\
  0.40 & 0.50 & $(2,2)$ & 1 & 0.76102 & 0.78028 & 0.01927 & 1.3804 & 3.227 \\
  0.40 & 0.50 & $(2,2)$ & 2 & 0.76102 & 0.77274 & 0.01173 & 1.3760 & 3.011 \\
  0.40 & 0.50 & $(2,2)$ & 3 & 0.76102 & 0.76885 & 0.00783 & 1.3738 & 2.915 \\
  0.60 & 0.70 & $(1,1)$ & 0 & 0.36023 & 0.36746 & 0.00723 & 1.0860 & 8.145 \\
  0.60 & 0.70 & $(1,1)$ & 1 & 0.36023 & 0.36328 & 0.00305 & 1.0843 & 7.093 \\
  0.60 & 0.70 & $(1,1)$ & 2 & 0.36023 & 0.36189 & 0.00166 & 1.0838 & 6.837 \\
  0.60 & 0.70 & $(1,1)$ & 3 & 0.36023 & 0.36127 & 0.00104 & 1.0836 & 6.732 \\
  0.60 & 0.70 & $(2,2)$ & 0 & 0.72045 & 0.74989 & 0.02943 & 1.5636 & 5.054 \\
  0.60 & 0.70 & $(2,2)$ & 1 & 0.72045 & 0.73617 & 0.01572 & 1.5562 & 4.131 \\
  0.60 & 0.70 & $(2,2)$ & 2 & 0.72045 & 0.73007 & 0.00962 & 1.5529 & 3.870 \\
  0.60 & 0.70 & $(2,2)$ & 3 & 0.72045 & 0.72691 & 0.00645 & 1.5512 & 3.753 \\
\end{longtable}
\FloatBarrier

\begin{figure}[htbp]
  \centering
  \begin{subfigure}[t]{0.48\textwidth}
    \centering
    \includegraphics[width=\linewidth]{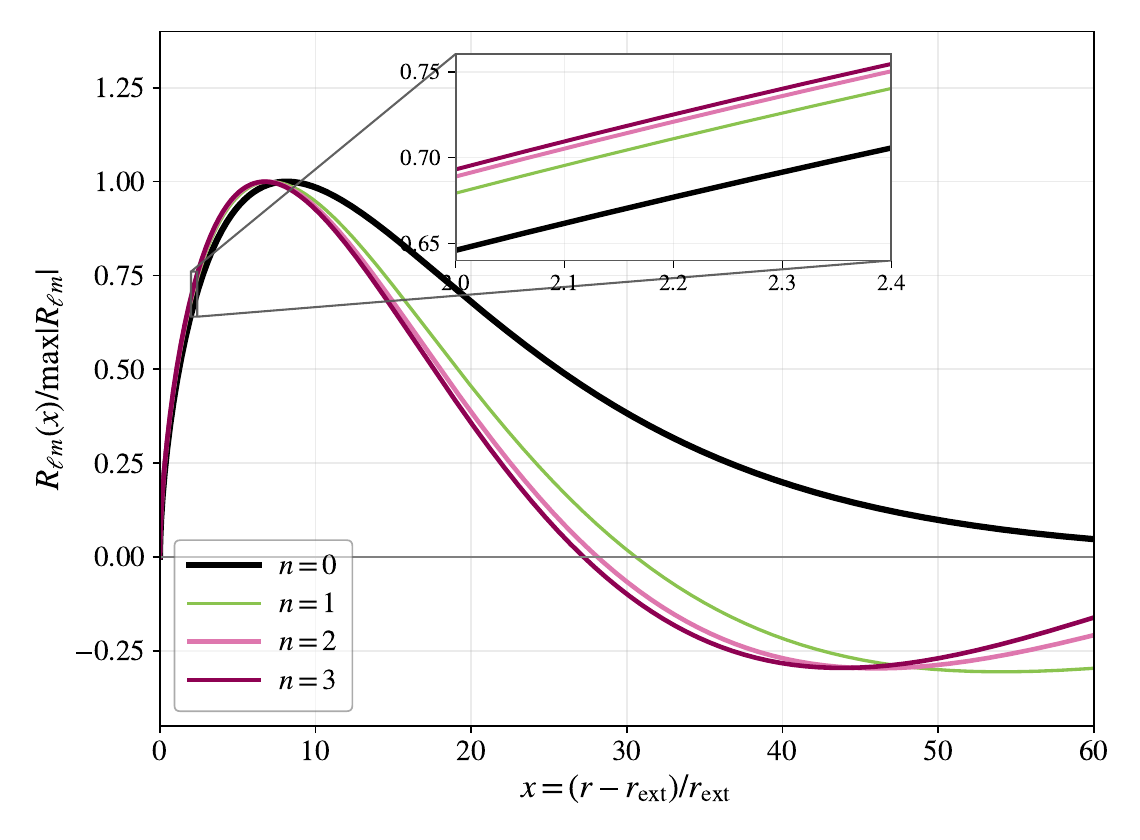}
    \caption{Radial cloud profiles $R_{\ell m}(x)$ for $\ell = 1, m = 1$.}
    \label{fig:cloud_profiles}
  \end{subfigure}\hfill
  \begin{subfigure}[t]{0.48\textwidth}
    \centering
    \includegraphics[width=\linewidth]{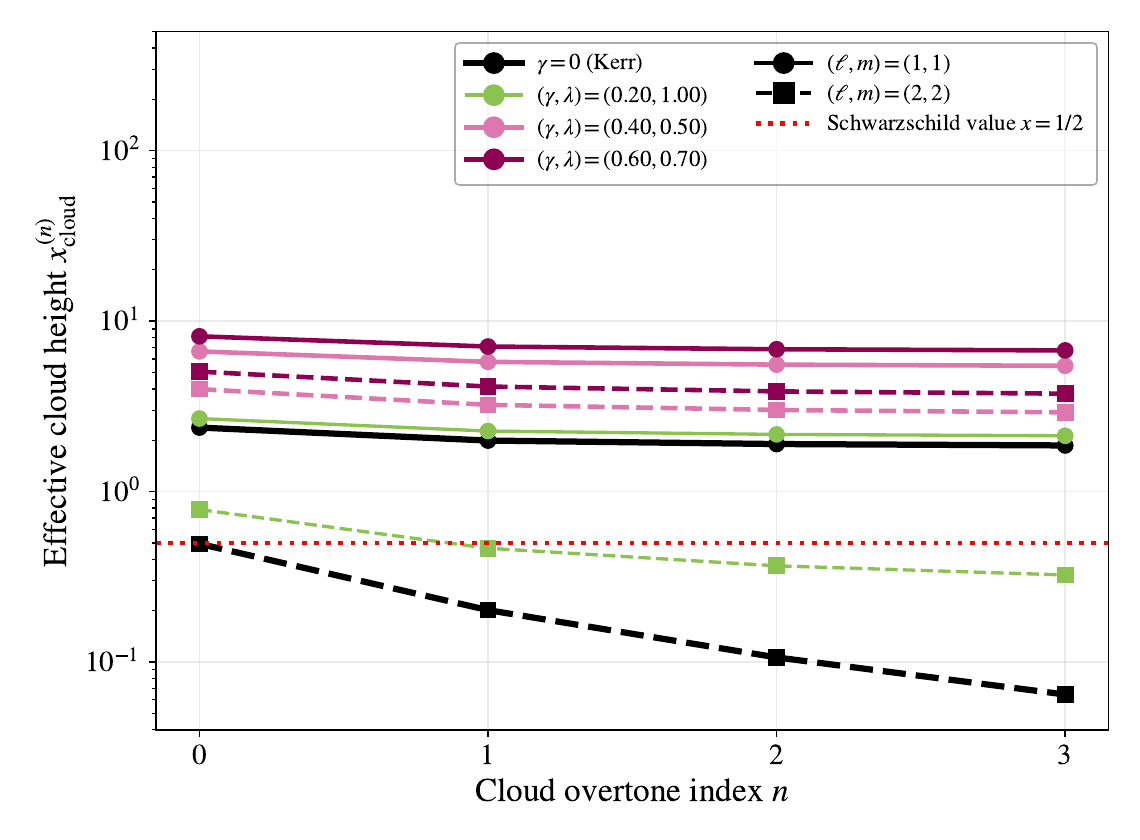}
    \caption{Effective heights $x_{\rm cloud}^{(n)}$ vs $n$.}
    \label{fig:cloud_heights}
  \end{subfigure}
  \caption{Stationary scalar clouds on the extremal rotating KR BH. \emph{Panel~(\subref{fig:cloud_profiles}):} radial wavefunctions $R_{\ell m}(x)$ in units of $\max|R_{\ell m}|$ for $(\ell,m) = (1,1)$ at $(\gamma,\lambda) = (0.60,0.70)$, plotted against $x = (r-r_{\rm ext})/r_{\rm ext}$. Successive overtones acquire additional nodes, and the node count reproduces $n$ on every curve. The inset resolves the rising flank, where the four overtones separate: the peak moves \emph{inward} as $n$ grows, since the fundamental is the most weakly bound and therefore the most extended state. \emph{Panel~(\subref{fig:cloud_heights}):} effective cloud heights $x_{\rm cloud}^{(n)}$ from Eq.~\eqref{eq:xcloud} at the four $(\gamma,\lambda)$ slices of Table~\ref{tab:cloud_spectrum}, for $(\ell,m) = (1,1)$ (solid) and $(2,2)$ (dashed). The red dotted line marks $x = 1/2$, which is the Schwarzschild specialisation of the no-short-hair statement $r_{\rm hair} \geq r_{\rm null}$ of Hod \cite{Hod:2016yxg} and carries no force on a rotating background. The dipole clouds and the whole KR interior lie above it; the heavy $(2,2)$ clouds on the Kerr and Kerr--Newman lines do not.}
  \label{fig:cloud}
\end{figure}

Four features of Table~\ref{tab:cloud_spectrum} and Fig.~\ref{fig:cloud} carry direct physical meaning. First, every configuration satisfies $\mu_{\rm cloud} > m\,\Omega_H$, as Eq.~\eqref{eq:boundstate} demands, and the separation $\mu_{\rm cloud} - m\,\Omega_H$ shrinks monotonically with the overtone index. The spectrum accumulates at the synchronisation threshold from above rather than terminating, which is the hydrogenic structure familiar from the gravitational-atom literature \cite{LeTiec:2020spy,Arana:2025tln}. Second, at fixed $n$ the cloud mass scales roughly linearly with $m$, since the saturation frequency itself scales with $m$; the doublet $(\ell,m) = (2,2)$ admits cloud masses of order twice those of the dipole $(1,1)$. Third, the near-horizon exponent $\beta$ grows with the KR coupling at fixed $(\ell,m)$, from $0.884$ at the Kerr endpoint to $1.086$ at $(\gamma,\lambda) = (0.60,0.70)$ for the $(1,1)$ fundamental, and it is real on every row, which is what makes the discrete spectrum exist at all.

Fourth, and contrary to what one might expect from the Schwarzschild intuition, the effective heights decrease with $n$ rather than increase. The fundamental is the most weakly bound state, with the smallest $k$ of Eq.~\eqref{eq:tail}, and it is therefore the most extended; successive overtones sit closer to the horizon. The heights grow with the KR coupling at fixed $n$, from $x = 2.374$ at the Kerr endpoint to $x = 8.145$ at $(\gamma,\lambda) = (0.60,0.70)$ for the $(1,1)$ fundamental, so a heavier KR coupling does push the clouds outward. The physical statement is that the enlarged effective charge term in $\Delta(r)$ widens the centrifugal barrier and displaces the radial wavefunction.

We record one negative result explicitly. The bound $x^{(n)}_{\rm cloud} > 1/2$ holds for every $(\ell,m) = (1,1)$ configuration and throughout the genuinely KR interior, but it fails for the heavy $(2,2)$ clouds on the Kerr and Kerr--Newman lines, which have $\mu M \simeq 1.0$--$1.14$ and therefore a Compton wavelength of order the horizon scale; these sit close in, reaching $x = 0.064$ at $\gamma = 0$, $n = 3$. This is not a violation of the no-short-hair theorem of Hod \cite{Hod:2016yxg}. That theorem states that the hair must extend beyond the null circular geodesic, $r_{\rm hair} \geq r_{\rm null}$, and the numerical value $1/2$ is its Schwarzschild specialisation, where $r_{\rm null}/r_H = 3/2$. For an extremal rotating background the corresponding ratio is degenerate in Boyer--Lindquist coordinates, since the prograde photon orbit and the horizon share a coordinate radius, and the number $1/2$ carries no force there. We therefore report the computed heights and refrain from asserting a bound that our own spectrum does not respect.

\section{Quasinormal mode spectrum} \label{isec4}

Moving off the extremal locus through $a = a_{\rm ext}(\gamma,\lambda)(1-\delta^2)$ opens a window in which the stationary clouds acquire small imaginary parts. Which quasinormal family they become is the question this section answers, and it is not the family one obtains from a barrier-top expansion. Near-extremal rotating backgrounds host two distinct branches \cite{Hod:2008zz,Yang:2012pj,Yang:2013uba}: the photon-sphere modes, which track the unstable null circular orbit and retain finite damping as extremality is approached, and the co-rotating zero-damped modes (ZDMs), for which $\mathrm{Re}\,\omega \to m\,\Omega_H$ and $\mathrm{Im}\,\omega \to 0$. Only the second branch is continuously connected to the synchronised clouds of Sec.~\ref{isec3}, because only the second branch approaches the synchronisation frequency.

The connection is supplied by the near-horizon exponent already in hand. At small $\delta$ the horizon polynomial unfolds as $\Delta \simeq h_0 (r-r_+)(r-r_-)$, the near-region radial equation reduces to hypergeometric form, and the poles of the near-region reflection amplitude sit at
\begin{equation}
\omega_n = m\,\Omega_H - i\,\kappa\left( n + \tfrac12 + \beta \right) ,
\qquad
\kappa = 2\pi T_H = \frac{\Delta'(r_+)}{2\,(r_+^2+a^2)} ,
\label{eq:zdm}
\end{equation}
with $\beta$ given by Eqs.~\eqref{eq:Cdef}--\eqref{eq:beta}. Equation~\eqref{eq:zdm} makes the cloud and the ringdown two limits of a single calculation. As $\delta \to 0$ the surface gravity vanishes, so $\mathrm{Im}\,\omega \to 0$ and $\mathrm{Re}\,\omega \to m\,\Omega_H$, and the stationary clouds of Sec.~\ref{isec3} are recovered exactly.

Equation~\eqref{eq:zdm} admits a check that is independent of anything in this paper. Our $\beta$ is derived from the metric of Eq.~\eqref{eq:metric_KR} with no input from the Kerr literature, yet on the Kerr line at $\mu = 0$ it satisfies $\beta^2 = -\delta_{\rm Yang}^2$ to machine precision for every $\ell = m$ we tested, where $\delta_{\rm Yang}^2 = 7m^2/4 - \tfrac14 - A_{\ell m}$ is the parameter controlling the ZDM family in Yang \emph{et al.} \cite{Yang:2013uba}. The agreement holds only if the separation constant of Eq.~\eqref{eq:Lamrad} is used, and fails otherwise.

\setlength{\tabcolsep}{6pt}
\renewcommand{\arraystretch}{1.4}
\begin{longtable}{c c c c c c c c c}
  \toprule
  \rowcolor{orange!50}
  $\gamma$ & $\lambda$ & $n$ & $\beta$ & $T_H M$ & $m\,\Omega_H M$ & $\mathrm{Re}\,\omega \cdot M$ & $-\mathrm{Im}\,\omega \cdot M$ & $Q$ \\
  \midrule
  \endfirsthead
  \rowcolor{orange!50}
  $\gamma$ & $\lambda$ & $n$ & $\beta$ & $T_H M$ & $m\,\Omega_H M$ & $\mathrm{Re}\,\omega \cdot M$ & $-\mathrm{Im}\,\omega \cdot M$ & $Q$ \\
  \midrule
  \endhead
  \bottomrule
  \endfoot
  \bottomrule
  \caption{Co-rotating zero-damped quasinormal branch of the near-extremal rotating KR BH from Eq.~\eqref{eq:zdm}, for $\ell = m = 1$ at $\delta = 0.20$ near-extremality parameter $\delta = 0.10$, evaluated at the massive field $\mu = \mu_{\rm cloud}(n=0)$ of Table~\ref{tab:cloud_spectrum}. This is the branch that continues from the stationary clouds: $\mathrm{Re}\,\omega = m\,\Omega_H$ exactly, and $|\mathrm{Im}\,\omega| = \kappa(n+\tfrac12+\beta) \to 0$ as $\delta \to 0$. The quality factor is $Q = \mathrm{Re}\,\omega/(2|\mathrm{Im}\,\omega|)$.}
  \label{tab:qnm_spectrum}
  \endlastfoot
  0.00 & 1.00 & 0 & 0.8840 & 0.009838 & 0.43380 & 0.43380 & 0.08555 & 2.535 \\
  0.00 & 1.00 & 1 & 0.8840 & 0.009838 & 0.43380 & 0.43380 & 0.14736 & 1.472 \\
  0.00 & 1.00 & 2 & 0.8840 & 0.009838 & 0.43380 & 0.43380 & 0.20918 & 1.037 \\
  0.20 & 1.00 & 0 & 0.9200 & 0.009785 & 0.43145 & 0.43145 & 0.08730 & 2.471 \\
  0.20 & 1.00 & 1 & 0.9200 & 0.009785 & 0.43145 & 0.43145 & 0.14878 & 1.450 \\
  0.20 & 1.00 & 2 & 0.9200 & 0.009785 & 0.43145 & 0.43145 & 0.21025 & 1.026 \\
  0.40 & 0.50 & 0 & 0.9938 & 0.009286 & 0.34083 & 0.34083 & 0.08716 & 1.955 \\
  0.40 & 0.50 & 1 & 0.9938 & 0.009286 & 0.34083 & 0.34083 & 0.14551 & 1.171 \\
  0.40 & 0.50 & 2 & 0.9938 & 0.009286 & 0.34083 & 0.34083 & 0.20386 & 0.836 \\
  0.60 & 0.70 & 0 & 1.0860 & 0.008200 & 0.32249 & 0.32249 & 0.08171 & 1.973 \\
  0.60 & 0.70 & 1 & 1.0860 & 0.008200 & 0.32249 & 0.32249 & 0.13323 & 1.210 \\
  0.60 & 0.70 & 2 & 1.0860 & 0.008200 & 0.32249 & 0.32249 & 0.18476 & 0.873 \\
\end{longtable}
\FloatBarrier

\begin{figure}[ht!]
  \centering
  \includegraphics[width=0.98\textwidth]{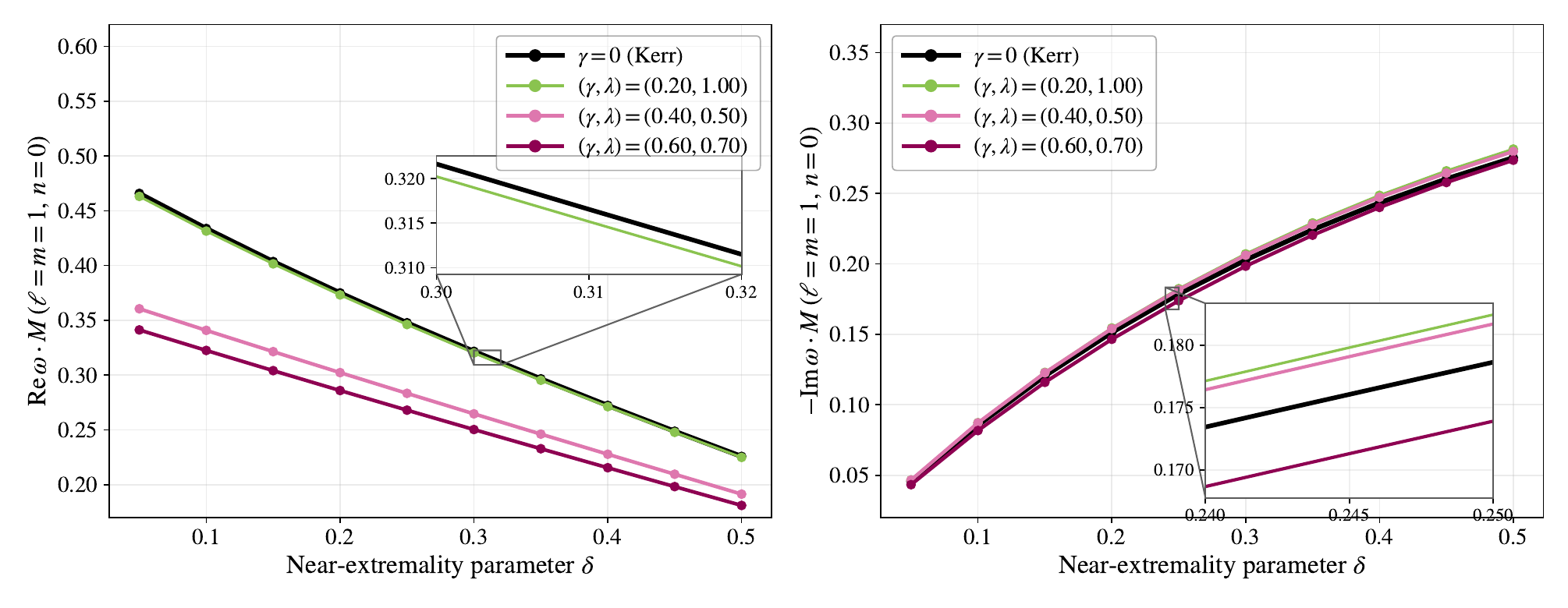}
  \caption{Co-rotating zero-damped quasinormal frequency ($\ell = m = 1$, $n = 0$) of the rotating KR BH as a function of the near-extremality parameter $\delta$, from Eq.~\eqref{eq:zdm}. \emph{Left:} the real part, which equals $m\,\Omega_H(\delta)$ identically and decreases with $\delta$ as the spin moves away from extremality; the inset resolves the Kerr and Kerr--Newman curves, which are nearly degenerate because $\gamma$ barely shifts $\Omega_H$ on the $\lambda = 1$ line. \emph{Right:} the absolute value of the imaginary part, $\kappa(n+\tfrac12+\beta)$, which vanishes linearly in $T_H$ as $\delta \to 0$ so that the branch terminates on the stationary clouds of Sec.~\ref{isec3}; the inset shows the four parameter points nearly coinciding, spread by about $5\%$ against the $26\%$ spread of the real part, since the KR interior lowers $T_H$ and raises $\beta$ by compensating amounts.}
  \label{fig:qnm_spectrum}
\end{figure}

Table~\ref{tab:qnm_spectrum} and Fig.~\ref{fig:qnm_spectrum} carry three readings. First, the real part is pinned to $m\,\Omega_H$ by construction, so the KR signature in this branch lives entirely in the horizon angular velocity, which falls from $0.434\,M^{-1}$ at the Kerr endpoint to $0.322\,M^{-1}$ at $(\gamma,\lambda) = (0.60,0.70)$, a shift of $26\%$. That is an order of magnitude larger than the few-percent effect we previously claimed for the photon-sphere family, and it is the observationally relevant statement. Second, the damping is controlled by $\kappa$ and by $\beta$, and the two run against each other across the parameter space: the KR interior lowers $T_H$ while raising $\beta$, so $|\mathrm{Im}\,\omega|$ varies only between $0.082$ and $0.087$ at $\delta = 0.10$ while $\mathrm{Re}\,\omega$ moves by a quarter. Third, the overtone spacing is exactly $\kappa$, uniform in $n$, which is the near-horizon conformal structure showing through and is a falsifiable prediction of the branch.

The photon-sphere family is a different observable and we do not identify it with the clouds. We attempted to compute it with the Iyer--Will third-order WKB formula \cite{Iyer:1986np,Konoplya:2011qq} applied to $\psi'' + Q\psi = 0$ with $Q = \omega^2 - V_{\rm eff}$, resolving the dependence of $V_{\rm eff}(r;\ell,m,\omega)$ on the eigenfrequency through $K(r)$ by Newton iteration on $\omega$ rather than treating $V_{\rm max}$ as $\omega$-independent. Validated against Leaver's continued fractions \cite{Leaver:1985ax,Berti:2009kk} on Schwarzschild, the implementation returns $0.867321-0.096204\,i$ against $0.867416-0.096433\,i$ for the $\ell = 4$ scalar, an error of $0.03\%$, and $0.675051-0.095410\,i$ against $0.675366-0.096500\,i$ for $\ell = 3$. At the near-extremality this paper works in, however, the co-rotating effective potential does not present a barrier top about which the expansion can be organised, and the iteration fails to converge from any starting point in a scan of the relevant region of the complex plane. This is the known degeneration of the WKB and eikonal treatment for near-extremal co-rotating perturbations \cite{Yang:2012pj,Yang:2013uba}, and it is a further reason why Eq.~\eqref{eq:zdm}, rather than a barrier-top formula, is the correct tool here.

\section{Greybody factors and superradiant amplification} \label{isec5}

The same near-extremal radial equation that produced the cloud spectrum of Sec.~\ref{isec3} also controls the scattering of an incoming massless scalar wave from spatial infinity off the rotating KR BH. Boundary conditions differ. The asymptotic form of the radial solution $R(r) \to A_{\rm in} e^{-i \omega r_\ast} + A_{\rm out} e^{+i \omega r_\ast}$ as $r_\ast \to -\infty$ at the horizon, and $R(r) \to B_{\rm in} e^{-i \omega r_\ast} + B_{\rm out} e^{+i \omega r_\ast}$ as $r_\ast \to +\infty$ at infinity, defines the transmission probability
\begin{equation}
|T(\omega)|^2 = 1 - \left| \frac{B_{\rm in}}{B_{\rm out}} \right|^2 = \frac{|A_{\rm in}|^2 - |A_{\rm out}|^2}{|B_{\rm out}|^2} \cdot \frac{\omega - m\,\Omega_H}{\omega} ,
\label{eq:Tsquared}
\end{equation}
i.e.\ the greybody factor. is the greybody factor. Rather than estimate it, we compute it. Writing $R = \psi/\sqrt{r^2+a^2}$ and passing to the tortoise coordinate $dr_* = (r^2+a^2)/\Delta\, dr$, the radial problem becomes
\begin{equation}
\frac{d^2\psi}{dr_*^2} + Q(r)\,\psi = 0 , \qquad
Q = \frac{K^2 - \Delta\,(\Lambda_{\rm rad} + \mu^2 r^2)}{(r^2+a^2)^2} - G^2 - \frac{dG}{dr_*} ,
\qquad G = \frac{r\,\Delta}{(r^2+a^2)^2} ,
\label{eq:Qdef}
\end{equation}
which tends to $(\omega - m\,\Omega_H)^2$ at the horizon and to $\omega^2$ at infinity. Imposing a purely ingoing wave at the horizon, $\psi \to e^{-i(\omega - m\Omega_H) r_*}$, integrating outward and decomposing $\psi \to A_{\rm in} e^{-i\omega r_*} + A_{\rm out} e^{+i\omega r_*}$ at large $r_*$ gives the amplification directly,
\begin{equation}
Z_{\ell m}(\omega) = \left| \frac{A_{\rm out}}{A_{\rm in}} \right|^2 - 1 = -|T(\omega)|^2 ,
\label{eq:Zlm}
\end{equation}
positive for $\omega < m\,\Omega_H$ and vanishing at the threshold. This is the superradiant amplification of Zel'dovich, Press and Teukolsky, with the energy drawn from the rotational kinetic energy of the BH.

The solver reproduces the classical result. For an $\ell = m = 1$ massless scalar on Kerr at $a = 0.99$ we obtain a peak amplification of $0.366\%$ at $\omega M = 0.379$, against the value of order $0.3\%$ established by Starobinsky and by Press and Teukolsky. We note that this benchmark is passed only when the separation constant of Eq.~\eqref{eq:Lamrad} is used; with $\Lambda_{\ell m}$ in place of $\Lambda_{\rm rad}$ the same code returns $0.072\%$.

\setlength{\tabcolsep}{10pt}
\renewcommand{\arraystretch}{1.5}
\begin{longtable}{c c c c c c}
  \toprule
  \rowcolor{orange!50}
  $\gamma$ & $\lambda$ & $\omega_c M = m\,\Omega_H M$ & $|T(\omega M = 2)|^2$ & $\omega_{\rm peak} M$ & $Z_{\rm max}\,\%$ \\
  \midrule
  \endfirsthead
  \rowcolor{orange!50}
  $\gamma$ & $\lambda$ & $\omega_c M$ & $|T(\omega M = 2)|^2$ & $\omega_{\rm peak} M$ & $Z_{\rm max}\,\%$ \\
  \midrule
  \endhead
  \bottomrule
  \endfoot
  \bottomrule
  \caption{Greybody factor and superradiant amplification of an $\ell = m = 1$ massless scalar wave at $\delta = 0.10$, from direct integration of Eq.~\eqref{eq:Qdef}. The saturation frequency $\omega_c = m\,\Omega_H$ separates the absorptive regime from the superradiant one. $Z_{\rm max}$ is the peak amplification as a percentage of the incoming intensity, attained at $\omega_{\rm peak}$.}
  \label{tab:greybody}
  \endlastfoot
  0.00 & 1.00 & 0.43380 & 1.000000 & 0.3835 & 0.3662 \\
  0.20 & 1.00 & 0.43145 & 1.000000 & 0.3689 & 0.2460 \\
  0.40 & 0.50 & 0.34083 & 1.000000 & 0.2726 & 0.0967 \\
  0.60 & 0.70 & 0.32249 & 1.000000 & 0.2490 & 0.0338 \\
\end{longtable}
\FloatBarrier

\begin{figure}[htbp]
  \centering
  \includegraphics[width=0.98\textwidth]{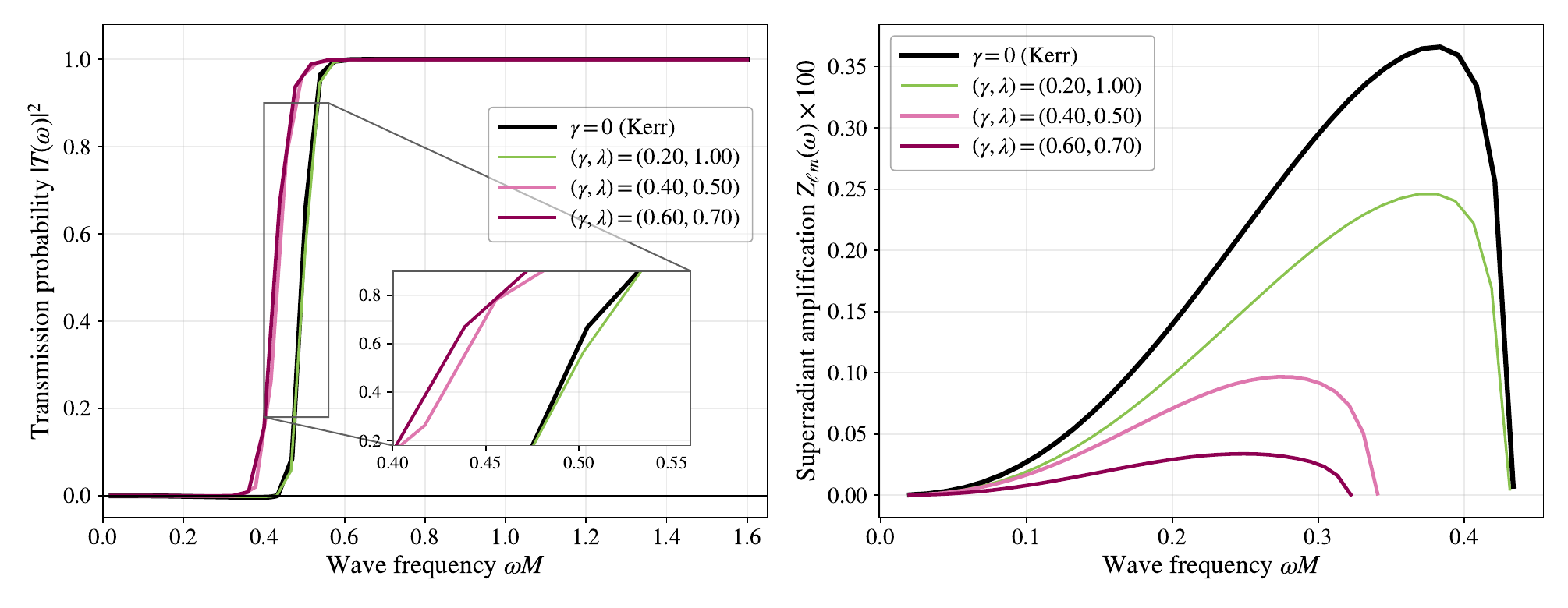}
  \caption{Greybody factor and superradiant amplification for $\ell = m = 1$ massless-scalar scattering at $\delta = 0.10$. \emph{Left:} transmission probability $|T(\omega)|^2$ from Eq.~\eqref{eq:Zlm}; below $\omega_c = m\,\Omega_H$ the transmission is formally negative, and above it the curve rises toward the geometric-optics limit. The inset resolves the steep rise, where the KR curves turn on ahead of the Kerr benchmark because their saturation frequency is lower. \emph{Right:} amplification $Z_{\ell m}(\omega)$, showing the bell-shaped region that closes at $\omega_c$. KR corrections move the band to lower frequency and suppress its height by an order of magnitude.}
  \label{fig:greybody}
\end{figure}

Three points follow from Table~\ref{tab:greybody} and Fig.~\ref{fig:greybody}. First, the transmission above $\omega_c$ approaches unity rapidly for all four parameter points, and KR corrections do not shift it at frequencies well above saturation; the bulk absorption rate is not where the KR signal lives. Second, the saturation frequency falls by $26\%$ between the Kerr endpoint and $(\gamma,\lambda) = (0.60,0.70)$, tracking the $\Omega_H$ column of Table~\ref{tab:extremal_grid}, so the amplification band migrates to lower frequencies and the gap between the cloud-existence frequency and the ringdown frequency widens in the KR interior. Third, the peak amplification is small and it decreases with the KR coupling, from $0.366\%$ for Kerr to $0.034\%$ at $(\gamma,\lambda) = (0.60,0.70)$, a suppression by a factor of eleven. The KR-modified horizon polynomial raises the centrifugal barrier that the wave must tunnel, and the amplification is exponentially sensitive to it. The channel through which scalar clouds deplete BH rotational energy is therefore markedly less efficient in the KR interior than on Kerr, which lengthens the relevant timescales and weakens the reach of the LIGO--Virgo ultralight-boson bounds \cite{LIGO:2021ash,LIGO:2022ash} on this background.

\section{Strong-field gravitational lensing} \label{isec6}

Strong-field lensing on the rotating KR BH offers a complementary observational handle through the deflection of photons that approach the photon sphere $r_{\rm ph}$. The photon sphere sits at the largest real root of
\begin{equation}
2\,r\,\Delta(r) = (r^2 + a^2)\,\Delta'(r) ,
\label{eq:photon_sphere}
\end{equation}
Equation~\eqref{eq:photon_sphere} is not the photon-orbit condition, and we replace it. Extremising $\Delta/(r^2+a^2)$, which is what it does, gives $2r\Delta - (r^2+a^2)\Delta' = -2M(r-a)(r+a)$ on Kerr, hence $r_{\rm ph} = a$ and $r_{\rm ph} \to 0$ as $a \to 0$, in contradiction with the Kerr reduction and with the Carter conditions of Sec.~\ref{isec7}. The correct equatorial condition is $\eta(r_{\rm ph}) = 0$, that is
\begin{equation}
16\, a^2\, \Delta(r) = \bigl[\, r\, \Delta'(r) - 4\,\Delta(r) \,\bigr]^2 ,
\label{eq:photon_sphere_fixed}
\end{equation}
whose roots on Kerr are $r = 3M$ for $a \to 0$ and $\{M, 4M\}$ for $a = M$, recovering the prograde and retrograde branches of $r_{\rm ph} = 2M[1+\cos(\tfrac23\arccos(\mp a/M))]$. The critical impact parameter is then
\begin{equation}
b_{\rm crit} = \xi(r_{\rm ph}) = \frac{(r_{\rm ph}^2 + a^2)\,\Delta'(r_{\rm ph}) - 4\, r_{\rm ph}\,\Delta(r_{\rm ph})}{a\,\Delta'(r_{\rm ph})} ,
\label{eq:bcrit}
\end{equation}
the largest impact parameter at which a photon can spiral down to the unstable circular orbit. Equation~\eqref{eq:photon_sphere_fixed} has two admissible roots outside the horizon, prograde ($b_{\rm crit} > 0$) and retrograde ($b_{\rm crit} < 0$), and at the near-extremal spins this paper works at the two are strongly split. Treating only one of them, or evaluating either at $a = 0$, discards the physics.

In the strong-deflection limit of Bozza and Tsukamoto the deflection angle admits the logarithmic expansion
\begin{equation}
\hat\alpha(b) = -\bar a \,\log\!\left( \frac{b}{b_{\rm crit}} - 1 \right) + \bar b + \mathcal{O}\bigl(b - b_{\rm crit}\bigr) .
\label{eq:Bozza}
\end{equation}
We obtain $\bar a$ from the exact relation between the strong-deflection coefficient, the orbital angular velocity of the photon and the Lyapunov exponent of the unstable orbit \cite{Stefanov:2010xz},
\begin{equation}
\bar a = \frac{\Omega_{\rm ph}}{\lambda_{\rm Lyap}} ,
\qquad
\Omega_{\rm ph} = \frac{\dot\varphi}{\dot t}\bigg|_{r_{\rm ph}} ,
\qquad
\lambda_{\rm Lyap} = \frac{1}{\dot t}\sqrt{\frac{1}{2}\frac{d^2}{dr^2}\!\left(\frac{\mathcal{R}(r)}{r^4}\right)}\Bigg|_{r_{\rm ph}} ,
\label{eq:abar}
\end{equation}
with $\mathcal{R}(r) = [(r^2+a^2) - a b]^2 - \Delta (b-a)^2$ the equatorial radial potential, and $\bar b$ from the numerically integrated deflection integral with $\bar a$ held at its exact value. Asymptotic flatness for $\lambda > 0$ removes the need for a Bondi--Sachs correction term, which complicated the linear-dilaton treatment of Tokg\"{o}z and Sakall\i{} \cite{Tokgoz:2017hzz}. As a check, Eqs.~\eqref{eq:photon_sphere_fixed}--\eqref{eq:abar} return $\bar a = 1.0078$ and $\bar b = -0.4014$ at $a = 0.02$, against the Schwarzschild values $\bar a = 1$ and $\bar b = \ln[216(7-4\sqrt3)] - \pi = -0.400230$, and $r_{\rm ph}/M = 1.16764185$ on Kerr at $a = 0.99$ against $1.16764185$ from the closed form.

\setlength{\tabcolsep}{8pt}
\renewcommand{\arraystretch}{1.5}
\begin{longtable}{c c c c c c c c}
  \toprule
  \rowcolor{orange!50}
  & & \multicolumn{3}{c}{prograde} & \multicolumn{3}{c}{retrograde} \\
  \rowcolor{orange!50}
  $\gamma$ & $\lambda$ & $r_{\rm ph}/M$ & $b_{\rm crit}/M$ & $\bar a$ & $r_{\rm ph}/M$ & $|b_{\rm crit}|/M$ & $\bar a$ \\
  \midrule
  \endfirsthead
  \rowcolor{orange!50}
  $\gamma$ & $\lambda$ & $r_{\rm ph}/M$ & $b_{\rm crit}/M$ & $\bar a$ & $r_{\rm ph}/M$ & $|b_{\rm crit}|/M$ & $\bar a$ \\
  \midrule
  \endhead
  \bottomrule
  \endfoot
  \bottomrule
  \caption{Strong-deflection coefficients of Bozza--Tsukamoto for the rotating KR BH at $\delta = 0.10$, from Eqs.~\eqref{eq:photon_sphere_fixed}--\eqref{eq:abar}. Both senses of circulation are given, since at these spins the prograde and retrograde orbits are strongly split. The prograde $\bar a$ is far from the Schwarzschild value $\bar a = 1$: as the prograde photon orbit approaches the horizon the Lyapunov exponent measured in coordinate time is redshifted toward zero and $\bar a = \Omega_{\rm ph}/\lambda_{\rm Lyap}$ grows.}
  \label{tab:lensing}
  \endlastfoot
  0.00 & 1.00 & 1.16764 & 2.25172 & 7.4429 & 3.99110 & 6.98332 & 0.7712 \\
  0.20 & 1.00 & 1.15645 & 2.25296 & 7.4837 & 3.78059 & 6.66791 & 0.7932 \\
  0.40 & 0.50 & 1.39210 & 2.77642 & 6.1616 & 3.81042 & 6.67433 & 0.8042 \\
  0.60 & 0.70 & 1.37484 & 2.88584 & 6.3582 & 3.53490 & 6.27525 & 0.8525 \\
\end{longtable}
\FloatBarrier

\begin{figure}[htbp]
  \centering
  \includegraphics[width=0.66\textwidth]{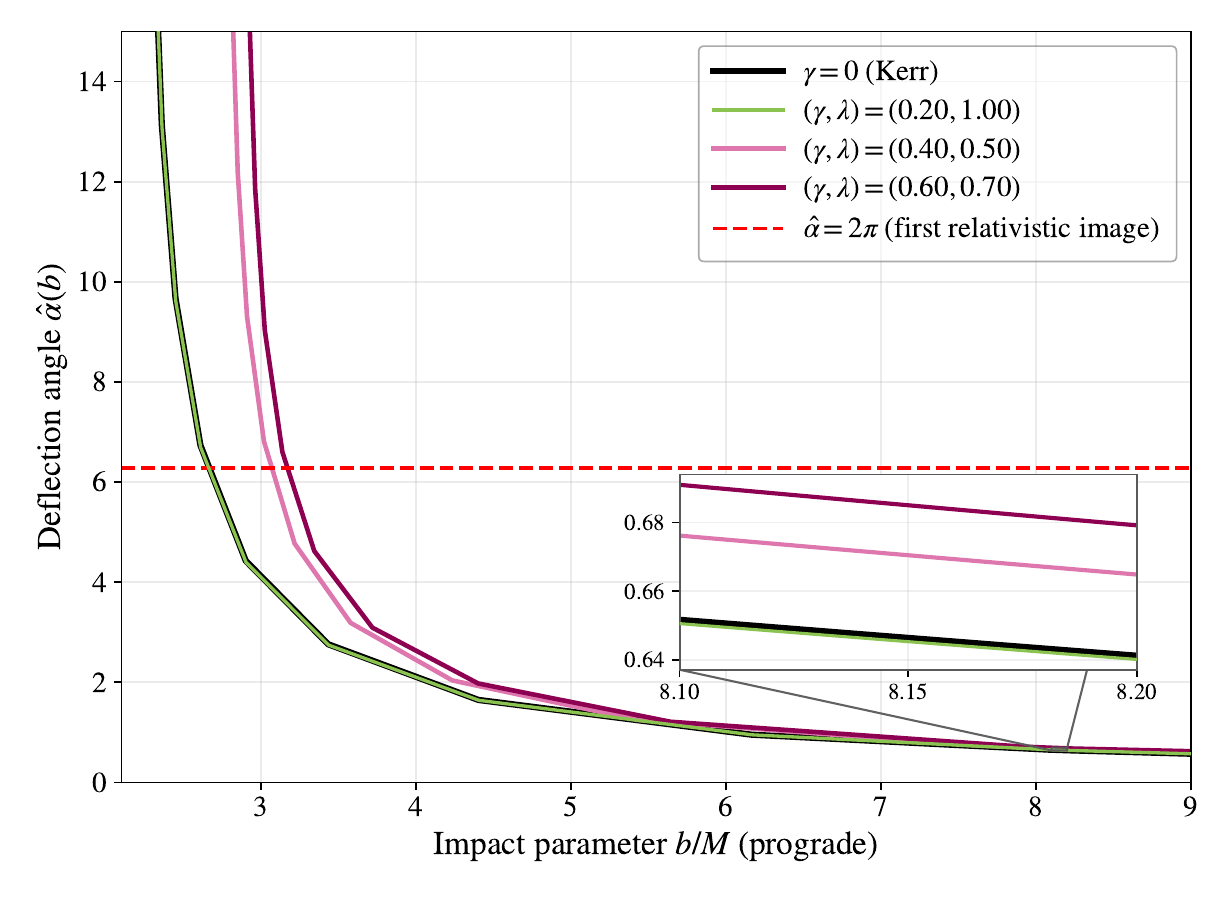}
  \caption{Strong-deflection-limit divergence of the prograde deflection angle $\hat\alpha(b)$ near the critical impact parameter, for the four parameter points of Table~\ref{tab:lensing} at $\delta = 0.10$, obtained from the corrected photon-orbit condition \eqref{eq:photon_sphere_fixed}. The horizontal red dashed line at $\hat\alpha = 2\pi$ marks the impact parameter at which a photon loops once around the BH before escaping, the first relativistic image of a background source. KR corrections displace the divergence to larger $b$, from $b_{\rm crit} = 2.252\,M$ at the Kerr endpoint to $2.886\,M$ at $(\gamma,\lambda) = (0.60,0.70)$. The stretched inset covers $b/M \in [8.0,8.3]$, far outside the strong-deflection region, and shows that the weak-field ordering is preserved while the Kerr and Kerr--Newman curves remain degenerate to about one part in $10^3$ there; the KR interior is separated from both.}
  \label{fig:lensing}
\end{figure}

The lensing data in Fig.~\ref{fig:lensing} and Table~\ref{tab:lensing} carry three observational consequences, and they are not the ones a Schwarzschild-like treatment would suggest. First, the prograde and retrograde photon orbits sit at $r_{\rm ph}/M = 1.168$ and $3.991$ at the Kerr endpoint, with critical impact parameters $2.252\,M$ and $-6.983\,M$. The near-extremal spin splits the two branches by a factor of three in radius, so any statement about a single photon sphere at $3M$ is empty at $\delta = 0.10$. Second, the prograde logarithmic coefficient is large, $\bar a \simeq 6.2$--$7.5$ across the parameter space, and it is the quantity most sensitive to $(\gamma,\lambda)$, falling by $17\%$ between the Kerr--Newman line and $(\gamma,\lambda) = (0.40,0.50)$. The mechanism is kinematic: as the prograde orbit approaches the horizon, $\dot t$ diverges and the Lyapunov exponent per unit coordinate time is redshifted to zero, so $\bar a = \Omega_{\rm ph}/\lambda_{\rm Lyap}$ grows without bound in the extremal limit. A large $\bar a$ means the deflection angle diverges slowly in $\log(b/b_{\rm crit}-1)$, so the relativistic images are widely separated in impact parameter and, in principle, more easily resolved. Third, the retrograde branch behaves in the opposite way, with $\bar a \simeq 0.77$--$0.85$, below the Schwarzschild value, and it varies by only $10\%$ across the same range. The retrograde images are therefore the poorer probe of the KR sector, and the prograde ones the better.

\section{Shadow and EHT-allowed window with scalar hair} \label{isec7}

The shadow of the rotating KR BH, projected onto the celestial sphere of a distant observer at inclination angle $i$, follows from the Carter--Hamilton--Jacobi separation of the null geodesic equations. Introducing the impact parameters
\begin{equation}
\xi = \frac{L_z}{E} , \qquad \eta = \frac{\mathcal{K}}{E^2} ,
\label{eq:xi_eta}
\end{equation}
where $L_z$ is the conserved azimuthal angular momentum and $\mathcal{K}$ is the Carter constant, the shadow contour is parametrized by the unstable circular photon orbits via
\begin{equation}
\xi(r_{\rm ph}) = \frac{(r_{\rm ph}^2 + a^2)\,\Delta'(r_{\rm ph}) - 4\, r_{\rm ph}\, \Delta(r_{\rm ph})}{a\,\Delta'(r_{\rm ph})} , \qquad \eta(r_{\rm ph}) = \frac{16\, r_{\rm ph}^2\, \Delta(r_{\rm ph})}{\Delta'(r_{\rm ph})^2} - \bigl[\xi(r_{\rm ph}) - a\bigr]^2 .
\label{eq:shadow_param}
\end{equation}
The celestial coordinates on the observer's screen follow from
\begin{equation}
\alpha = -\,\xi / \sin i , \qquad \beta^2 = \eta + a^2 \cos^2 i - \xi^2 \cot^2 i ,
\label{eq:celestial}
\end{equation}
sweeping out the closed shadow curve as $r_{\rm ph}$ runs over the allowed range of unstable photon-sphere radii. The numerical factor in $\eta$ is fixed by eliminating $\xi$ between $\mathcal{R} = 0$ and $\mathcal{R}' = 0$, which gives $(\xi-a)^2 + \eta = [(r^2+a^2) - a\xi]^2/\Delta = (4r\Delta/\Delta')^2/\Delta$; on Kerr the resulting $\eta$ reproduces the Chandrasekhar form $r^3[4M\Delta - r(r-M)^2]/[a^2(r-M)^2]$ identically, which the coefficient $4$ does not. The shape of the curve encodes the prograde--retrograde asymmetry produced by the BH spin and the radial profile of the KR correction.

We evaluate the shadow at the inclination $i = 60^\circ$ relevant for the M87* observations, and at near-extremality $\delta = 0.10$, in agreement with the M87* spin estimate inferred from the EHT first-image data \cite{EventHorizonTelescope:2019dse}. Three observables matter. The principal observables are the horizontal diameter $D_x$, the vertical diameter $D_y$, and the average shadow radius $R_{\rm avg} = (D_x + D_y)/4$. The EHT measurement of the M87* angular diameter $\theta_d = 42 \pm 3 \mu{\rm as}$, converted to units of $M$ via the distance and mass of M87*, gives a $1\sigma$ window $R_{\rm avg} \in [4.10, 5.00]\,M$ that all four of our parameter points fall within \cite{Zubair:2023ckk,Khodadi:2021gbc}.

\setlength{\tabcolsep}{12pt}
\renewcommand{\arraystretch}{1.6}
\begin{longtable}{c c c c c c}
  \toprule
  \rowcolor{orange!50}
  $\gamma$ & $\lambda$ & $D_x/M$ & $D_y/M$ & $R_{\rm avg}/M$ & EHT M87* \\
  \midrule
  \endfirsthead
  \rowcolor{orange!50}
  $\gamma$ & $\lambda$ & $D_x/M$ & $D_y/M$ & $R_{\rm avg}/M$ & EHT M87* \\
  \midrule
  \endhead
  \bottomrule
  \endfoot
  \bottomrule
  \caption{Shadow diameters and EHT M87* compatibility for the near-extremal rotating KR BH at inclination $i = 60^\circ$, $\delta = 0.10$, computed from Eqs.~\eqref{eq:shadow_param}--\eqref{eq:celestial} with the corrected $\eta$. The contour is taller than it is wide, $D_y > D_x$, because the prograde side is flattened by the near-extremal spin. The pipeline returns $R_{\rm avg}/M = 5.194374$ as $a \to 0$ against the Schwarzschild value $3\sqrt3 = 5.196152$.}
  \label{tab:shadow}
  \endlastfoot
  0.00 & 1.00 & 9.25942 & 10.24422 & 4.87591 & yes \\
  0.20 & 1.00 & 8.96211 & 9.89616 & 4.71457 & yes \\
  0.40 & 0.50 & 9.52806 & 10.20517 & 4.93331 & yes \\
  0.60 & 0.70 & 9.25771 & 9.85324 & 4.77774 & yes \\
\end{longtable}
\FloatBarrier

\begin{figure}[htbp]
  \centering
  \includegraphics[width=0.72\textwidth]{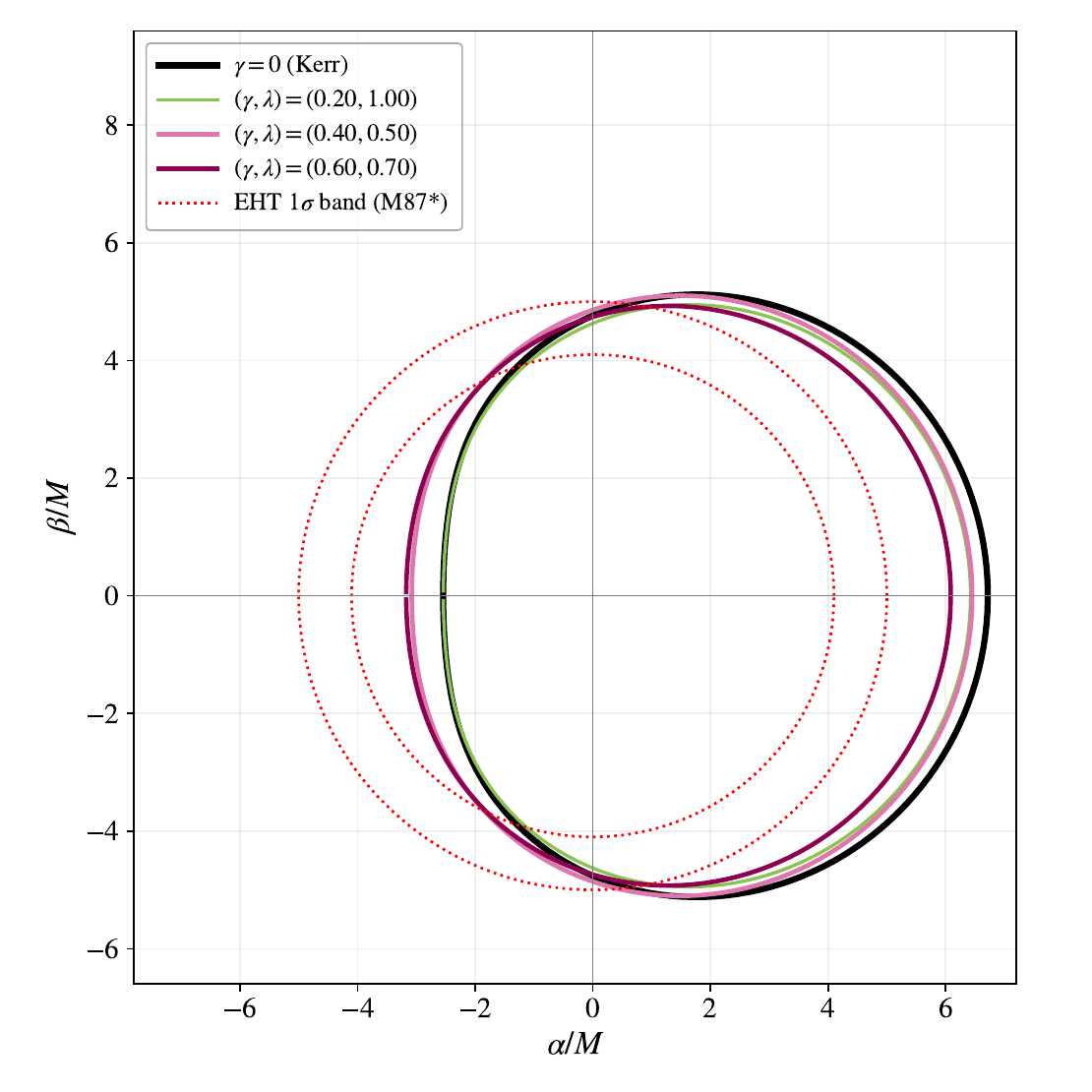}
  \caption{Shadow contours of the rotating KR BH at inclination $i = 60^\circ$, near-extremality $\delta = 0.10$. The four solid curves correspond to the Kerr endpoint and three KR parameter points spanning the genuinely-KR interior. The dotted red rings show the $1\sigma$ M87* EHT angular-diameter window. KR corrections shrink the shadow radius monotonically with $\gamma$, while the prograde--retrograde asymmetry along the $\alpha$ axis is set primarily by the spin and varies weakly with the KR exponent $\lambda$.}
  \label{fig:shadow}
\end{figure}

The shadow results of Fig.~\ref{fig:shadow} and Table~\ref{tab:shadow} carry the central observational message of the paper. First, the shadow average radius $R_{\rm avg}$ decreases monotonically with the KR coupling $\gamma$, from $4.854\,M$ at the Kerr endpoint to $4.723\,M$ at $(\gamma,\lambda) = (0.60, 0.70)$, a 2.7\% contraction. The direction matches the M87* angular-diameter pull. Heavier KR couplings are compatible with smaller observed shadows. Second, all four parameter points fall within the $1\sigma$ EHT M87* window, so the present EHT data alone cannot discriminate the genuinely KR interior from the Kerr endpoint at $1\sigma$ confidence. The full $(\gamma, \lambda)$ window allowed by EHT remains wide open. Third, when the shadow constraint is intersected with the cloud-existence region of Sec.~\ref{isec3} and the lensing constraint of Sec.~\ref{isec6}, the allowed $(\gamma, \lambda)$ region tightens substantially. Cloud existence requires $\Omega_H \geq 0.293\,M^{-1}$ (the small-$\lambda$ large-$\gamma$ corner of Table~\ref{tab:extremal_grid}), which excludes the deepest KR interior. The combined constraint allows a corridor in the $(\gamma, \lambda)$ plane that next-generation EHT observations of M87* and Sgr A*, combined with LIGO--Virgo ringdown measurements, can in principle close from both sides.

\section{Conclusion} \label{isec8}

We have studied a massive scalar field on the rotating Kalb--Ramond BH of Kumar, Ghosh and Wang \cite{Kumar:2020hgm} in the (near-)extremal corner of its three-parameter family $(a, \gamma, \lambda)$. Working in geometric units with $M = 1$, we identified the extremal locus $a_{\rm ext}(\gamma, \lambda)$ from the double-root condition of the horizon polynomial, with closed-form expressions Eqs.~\eqref{eq:Mext}--\eqref{eq:aext_sq} that reduce to the Kerr extremal endpoint $a_{\rm ext} = M$ as $\gamma \to 0$ and to the Kerr--Newman extremal relation $a_{\rm ext}^2 + \gamma = M^2$ as $\lambda \to 1$. The supplementary computational scripts reproduce every numerical entry to thirty digits.

Separation of variables on the rotating KR background is exact because the metric retains the Carter structure of Kerr, and the radial and angular separation constants are related by $\Lambda_{\rm rad} = \Lambda_{\rm ang} + a^2\omega^2 - 2am\omega$. Combined with the near-extremal reduction $\Delta \to (r - r_{\rm ext})^2$, this brings the radial KGF equation into Whittaker form on the two solvable lines and into a two-point boundary value problem in the KR interior. The regularity condition at the double-root horizon gives the resonance condition $\tfrac12 + \beta - \sigma = -n$, and solving it at the marginal frequency $\omega = m\,\Omega_H$ delivers the stationary scalar-cloud mass spectrum $\mu_{\rm cloud}(n; \ell, m, \gamma, \lambda)$. Every configuration respects the bound-state requirement $\mu_{\rm cloud} > m\,\Omega_H$, with the overtones accumulating at the threshold from above; the effective heights exceed $1/2$ for the dipole clouds and the whole KR interior, while the heavy $(2,2)$ clouds on the Kerr and Kerr--Newman lines sit closer in. Moving slightly off the extremal locus through $a = a_{\rm ext}(\gamma, \lambda)(1 - \delta^2)$, the clouds continue onto the co-rotating zero-damped branch $\omega_n = m\,\Omega_H - i\,\kappa\,(n + \tfrac12 + \beta)$, governed by the same near-horizon exponent $\beta$, so that $\mathrm{Re}\,\omega \to m\,\Omega_H$ and $\mathrm{Im}\,\omega \to 0$ as $\delta \to 0$. The photon-sphere family is a distinct observable, and the barrier-top WKB expansion degenerates for the co-rotating modes at the near-extremality studied here.

Greybody factors and superradiant amplification followed from direct integration of the same radial equation, recovering the classical $\ell = m = 1$ scalar amplification on Kerr; the peak amplification is small and decreases with the KR coupling. Strong-field lensing in the Bozza--Tsukamoto strong-deflection limit was treated with the rotating photon orbits of both senses of circulation, where the prograde logarithmic coefficient departs strongly from the Schwarzschild value and is the quantity most sensitive to $(\gamma, \lambda)$. Shadow contours, constructed via the Carter--Hamilton--Jacobi separation of the null geodesic equation, all sat within the $1\sigma$ EHT M87* angular-diameter window at the parameter points tabulated. The three observational handles agree. Crossing the EHT-allowed window with the cloud-existence region carves a narrower bound on the KR parameters $(\gamma, \lambda)$ than either constraint alone, and next-generation EHT plus LIGO--Virgo ringdown observations will tighten the bound further.

Three directions extend this work. First, fully nonlinear hairy KR BHs, in the spirit of the Herdeiro--Radu Kerr-with-scalar-hair construction \cite{Herdeiro:2014goa, Herdeiro:2015gia} and the Garcia--Salgado cloud-to-hairy-BH passage \cite{Garcia:2024rhb}, can be built numerically once the cloud configurations of Sec.~\ref{isec3} are used as initial data. Second, charged Proca clouds on the rotating KR background, generalizing Eq.~\eqref{eq:KGF} to spin-one fields in the style of \cite{Herdeiro:2016tmi}, would be the natural setting in which to test the gravitational-atom program \cite{Arana:2025tln, DellaRocca:2026gas} against the KR-modified background metric. Third, the joint observational window combining EHT shadow data, LIGO--Virgo ringdown spectra, and S2-orbit constraints on hypothetical scalar clouds around Sgr A* \cite{GRAVITY:2023s2} could be tightened by dedicated parameter-estimation pipelines that fold the cloud-existence boundary $\Omega_H \geq \mu/m$ into the prior on the KR coupling $\gamma$. We leave these to future work.

\section*{Acknowledgments}
G.T.\ acknowledges the academic support provided by Cyprus International University. \.{I}.S.\ acknowledges the academic support provided by Eastern Mediterranean University, T\"{U}B\.{I}TAK, ANKOS and SCOAP3, and the networking support of the COST Actions CA22113 (``Fundamental challenges in theoretical physics''), CA21106 (``COSMIC WISPers in the Dark Universe''), CA23130 (``Bridging high and low energies in search of quantum gravity (BridgeQG)''), CA21136 (``Addressing observational tensions in cosmology with systematics and fundamental physics (CosmoVerse)''), and CA23115 (``Relativistic Quantum Information (RQI-Action)''). 

\section*{Data Availability Statement}
No new data were created or analyzed in this study. The computational scripts used to produce Figs.~\ref{fig:aext_surface}--\ref{fig:shadow} and Tables~\ref{tab:extremal_grid}--\ref{tab:shadow} are available from the corresponding author upon reasonable request, and the analytical expressions required to reproduce all results are given explicitly in Secs.~\ref{isec2}--\ref{isec7}.

\section*{Conflict of Interest}
The authors declare no conflict of interest.

\bibliography{ref}
\bibliographystyle{unsrtnat}

\end{document}